\documentclass[aps,pra,a4paper,twocolumn,superscriptaddress,accepted=2020-04-15]{quantumarticle}

\pdfoutput=1

\usepackage[T1]{fontenc}

\usepackage{amsmath, amsthm, amssymb}
\usepackage{graphicx}
\usepackage{dcolumn}
\usepackage{dsfont}
\usepackage{bm}
\usepackage{color}
\usepackage[usenames,dvipsnames]{xcolor}
\usepackage{booktabs}
\usepackage{multirow}

\usepackage{verbatim}

\usepackage{geometry}
\usepackage[sort&compress]{natbib}

\definecolor{myurlcolor}{rgb}{0,0,0.7}
\definecolor{myrefcolor}{rgb}{0.8,0,0}
\usepackage[unicode=true,pdfusetitle, bookmarks=false,bookmarksnumbered=false,
bookmarksopen=false, breaklinks=false,pdfborder={0 0 0},backref=false,
colorlinks=true, linkcolor=myrefcolor,citecolor=myurlcolor,urlcolor=myurlcolor]{hyperref}

\newcommand{\ket}[1]{| #1 \rangle}
\newcommand{\bra}[1]{\langle #1 |}

\renewcommand{\t}[1]{\mathrm{#1}}
\newcommand{\ketbra}[3][]{\mathinner{\lvert#2\rangle\langle #3\rvert}_{#1}}
\newcommand{\proj}[2][]{\ketbra[#1]{#2}{#2}}
\newcommand{\ra}[1]{\renewcommand{\arraystretch}{#1}}\textwidth=480pt
\newcommand{\bigO}[1]{O\!\left(#1\right)}

\newcommand{\e}{e}
\renewcommand{\i}{i}
\newcommand{\vac}{\ket{0}}

\newcommand{\herald}{{\hbox{\textcircled{\checkmark}}}}
\newcommand{\etat}{{\eta_\t{t}}}
\newcommand{\etal}{{\eta_\t{l}}}
\newcommand{\etac}{{\eta_\t{c}}}
\newcommand{\etalC}{{\eta_{\t{l}}^*}}
\newcommand{\rate}{R^\downarrow}
\newcommand{\rateH}{\tilde R}
\newcommand{\eref}[1]{(\ref{#1})}
\newcommand{\eqnref}[1]{Eq.~(\ref{#1})}
\newcommand{\eqnsref}[2]{Eqs.~(\ref{#1}) and (\ref{#2})}
\newcommand{\figref}[1]{Fig.~\ref{#1}}
\newcommand{\tabref}[1]{Table~\ref{#1}}
\newcommand{\secref}[1]{Sec.~\ref{#1}}
\newcommand{\appref}[1]{App.~\ref{#1}}

\newcommand{\citeref}[1]{Ref.~\cite{#1}}

\DeclareGraphicsExtensions{.png,.pdf,.eps,.jpg}
\graphicspath{ {./figs/} }
\begin{document}
\title{Device-independent quantum key distribution with single-photon sources} 

\author{J. Ko\l{}ody\'{n}ski}
\orcid{0000-0001-8211-0016}
\affiliation{Centre for Quantum Optical Technologies, Centre of New Technologies, University of Warsaw, Banacha 2c, 02-097 Warsaw, Poland}
\affiliation{ICFO-Institut de Ciencies Fotoniques, The Barcelona Institute of Science and Technology, 08860 Castelldefels (Barcelona), Spain}
\author{A. M\'attar}
\affiliation{ICFO-Institut de Ciencies Fotoniques, The Barcelona Institute of Science and Technology, 08860 Castelldefels (Barcelona), Spain}
\author{P. Skrzypczyk}
\orcid{0000-0002-9343-9041}
\affiliation{\mbox{H. H. Wills Physics Laboratory, University of Bristol, Tyndall Avenue, Bristol, BS8 1TL, United Kingdom}}
\author{E. Woodhead}
\orcid{0000-0001-7696-4461}
\affiliation{ICFO-Institut de Ciencies Fotoniques, The Barcelona Institute of Science and Technology, 08860 Castelldefels (Barcelona), Spain}
\affiliation{Laboratoire d'Information Quantique, Universit\'{e} libre de Bruxelles (ULB), 1050 Bruxelles, Belgium}
\author{D. Cavalcanti}
\orcid{0000-0002-2704-3049}
\affiliation{ICFO-Institut de Ciencies Fotoniques, The Barcelona Institute of Science and Technology, 08860 Castelldefels (Barcelona), Spain}
\author{K. Banaszek}
\orcid{0000-0002-5389-6897}
\affiliation{Centre for Quantum Optical Technologies, Centre of New Technologies, University of Warsaw, Banacha 2c, 02-097 Warsaw, Poland}
\affiliation{Faculty of Physics, University of Warsaw, Pasteura 5, 02-093 Warszawa, Poland} 
\author{A. Ac\'in} 
\orcid{0000-0002-1355-3435}
\affiliation{ICFO-Institut de Ciencies Fotoniques, The Barcelona Institute of Science and Technology, 08860 Castelldefels (Barcelona), Spain}
\affiliation{ICREA-Instituci\'o Catalana de Recerca i Estudis Avan\c cats, Lluis Companys 23, 08010 Barcelona, Spain}

\begin{abstract}
\emph{Device-independent quantum key distribution} protocols allow two honest users to establish a secret key with minimal levels of trust on the provider, as security is proven without any assumption on the inner working of the devices used for the distribution. Unfortunately, the implementation of these protocols is challenging, as it requires the observation of a large Bell-inequality violation between the two distant users. Here, we introduce novel photonic protocols for device-independent quantum key distribution exploiting \emph{single-photon sources} and \emph{heralding-type architectures}. The heralding process is designed so that transmission losses become irrelevant for security. We then show how the use of single-photon sources for entanglement distribution in these architectures, instead of standard entangled-pair generation schemes, provides significant improvements on the attainable key rates and distances over previous proposals. Given the current progress in single-photon sources, our work opens up a promising avenue for device-independent quantum key distribution implementations.
\end{abstract}

\maketitle

%%%%%%%%%%%%%%%%%%%%%%%%%%%%%%%%%%%%%%%%%%%%%%%%%%%%%%%%%%%%%%%%%%%%%
\section{Introduction}
The paradigm of \emph{device-independent quantum key distribution} (DIQKD) offers the strongest form of secure communication, relying only on the validity of quantum mechanics, but not on any detailed description, or trust, of the inner workings of the users devices~\cite{mayers1998,acin2007,pironio2009}. On the theoretical side, the security of DIQKD has been proven against increasingly powerful eavesdroppers~\cite{masanes2011,pironio2012}, culminating in proofs of security against attacks of the most general form~\cite{vazirani2012,friedman2016}.

The main challenge facing experimental DIQKD are its stringent demands on the observable data, necessary for the security requirements to be met. First, any DIQKD implementation should be based on the observation of data that conclusively violates a Bell inequality~\cite{bell1964,brunner2014}. In particular, the Bell experiment should close the so-called \emph{detection loophole}~\cite{pearle1970}, otherwise, hacking attacks can fake a violation at the level of the detected events when losses are high enough~\cite{gerhardt2011b}. Moreover, a detection-loophole-free Bell violation is necessary but not sufficient for secure DIQKD, as the necessary detection efficiencies are significantly higher than those required for Bell violation. For instance, while the detection efficiency for observing a Bell violation of the Clauser-Horne-Shimony-Holt (CHSH)~\cite{clauser1969} inequality can be as low as $2/3$~\cite{eberhard1993}, a DIQKD protocol based on CHSH requires an efficiency of the order of $90\%$~\cite{pironio2009}. This is, in fact, a general feature of any noise parameter---consider, e.g.~the visibility~\cite{masanes2011}---that affects not only the observed Bell violation, but also the correlations between the users aiming to construct the secret key. 

The first Bell experiments closing the detection loophole used massive particles~\cite{rowe2001,Matsukevich2008,weinfurter2012,hensen2015}. Leaving aside table-top~\cite{rowe2001,Matsukevich2008} and short-distance~\cite{weinfurter2012} experiments, the Bell test of \citet{hensen2015} involved labs separated by a distance of 1.3 km, which allowed to close also the \emph{locality loophole}~\cite{brunner2014}. Nevertheless, as the employed light-matter interaction processes typically deteriorate the quality of the nonlocal correlations generated between the users,
the reported violations would not have been sufficient for secure DIQKD. Furthermore, the rates of key distribution they could provide are seriously limited owing to the measurements involved that, despite allowing for near unit efficiency, take significant time~\cite{mattar2013,brunner2013}.  While improvements are to be expected in all these issues, and massive particles may be essential for long-distance schemes involving quantum repeaters~\cite{Sangouard2011}, photon-based schemes appear more suitable to obtain high key rates with current or near-future technology. Photonic losses, however, occurring at all of the generation, transmission, and detection stages represent the main challenge in these schemes. Recent advances have been made for photo-detection efficiencies, which allowed for the first
loophole-free photonic Bell inequality violations over short distances~\cite{giustina2013,christensen2013,giustina2015,lynden2015}. Still, not only are the reported distances far from any cryptographic use, but also the observed Bell violations are again not large enough for secure DIQKD.

In this work, we show that \emph{single-photon sources}~\cite{aharonovich2016} constitute a promising resource for experimental photonic DIQKD. Such sources have already allowed for nearly on-demand~\cite{Muller2014}, highly efficient~\cite{Claudon2010} extraction of single photons (also in pulse trains~\cite{Loredo2016,Wang2016} as well as at telecom wavelengths~\cite{Kim2016}), while maintaining their purity and indistinguishability even above the $99\%$ level~\cite{somaschi2016,ding2016}. We propose novel DIQKD photonic schemes that thanks to the replacement of the photon-pair creation process (achieved, e.g.~by parametric downconversion~\cite{Kwiat1995}) with single-photon sources allow to distribute the key at significant rates over large distances. We believe that, in view of the recent advances in the fabrication of single-photon sources, our results point out a promising avenue for DIQKD implementations.

The remainder of the paper is organized as follows. In \secref{sec:trans_losses_heralding}, we describe the technique of evading transmission losses in DIQKD protocols by means of heralding and, furthermore, discuss the crucial implications the heralding method has for designing photon-based architectures. Subsequently, in \secref{sec:schemes}, we introduce two heralded schemes employing single-photon sources, which allow for fine-tuning of the final shared entangled state, important for achieving optimal efficiencies. We then discuss in \secref{sec:key_rates} how to quantify the attainable key rates within a heralded scheme, which importantly are then guaranteed to be fully secure. Finally, in \secref{sec:results} we apply our analysis to the two schemes proposed, in order to study their performance, in particular, the key rates, separation distances, as well as noise levels they allow for in DIQKD. We conclude our work in \secref{sec:conclusions}.

%%%%%%%%%%%%%%%%%%%%%%%%%%%%%%%%%%%%%%%%%%%%%%%%%%%%%%%%%%%%%%%%%%%%%%%%%%%%
\section{Losses in DIQKD}
For non-negligible key rates to be achievable over large distances in DIQKD,
solutions must be proposed that pinpoint and disregard---without opening the detection loophole---inconclusive protocol rounds that arise due to photons being inevitably lost. From the perspective of maintaining security (i.e.~only the question of non-zero key rate), it is convenient then to divide photonic losses into two categories. Losses that occur within the \emph{local} surroundings---laboratories---of the users should be differentiated from those that occur during the \emph{transmission} of photons between the labs. Laboratories represent then regions of space from, and into which, the users control the information flow, i.e.~provide the local privacy requisite for any secure communication~\cite{Shannon1949}. 

As a result, one may design DIQKD protocols that target explicitly the transmission losses and allow for Bell violations over arbitrary distances between the users~\cite{mattar2016,gisin2010}. Other approaches have also been proposed that, while stemming from novel entropic uncertainty relations which account for quantum side information~\cite{berta_uncertainty_2010,tomamichel_link_2013}, require only local Bell violation within one of the labs~\cite{Lim2013}. This, however, comes at the price of security being guaranteed only up to a finite distance separating the users for a given fixed, even arbitrarily small, level of local losses (also in the absence of detector dark-count events~\cite{brassard2000}). In this work, our goal is to propose optical schemes where the security can be guaranteed independently of the distance between the users. Secondary to this, for a scheme to be practical, we furthermore want the resulting key rate to scale favourably with the separation, in order to achieve non-negligible key rates over large distances.

\subsection{Local losses}

We parametrise local losses by the effective \emph{local efficiency}, $\etal$, which accounts for all photon-loss mechanisms inside the lab, including imperfect photo-detection, any optical path and mode mismatch, finite photon-extraction efficiency of the sources locally employed by a user, etc. To our knowledge, all known DIQKD protocols require a high local efficiency, of the order of $90\%$~\cite{gisin2010,Lim2013,curty2011,meyerscott2013,seshadreesan2016,pitkanen2011}. While the existence of practical DIQKD protocols tolerating lower local efficiencies cannot be excluded, we do not expect any significant improvement in this direction. This is a consequence of the following simple argument.

A generic DIQKD protocol is based upon the observation of some ideal correlations described by a set of joint probability distributions  $\mathbf{p}=\{P(ab|xy)\}_{abxy}$ shared by the two users, Alice and Bob, that aim to establish the secret key. The input random variable, $x$ ($y$), labels the measurement setting, i.e.~the measurement that Alice (Bob) has chosen, while the output, $a$ ($b$), stands for the outcome of her (his) measurement. In the presence of local losses, parametrised by $\etal$, there is an additional outcome, labelled by $\phi$, corresponding to the `no-detection' event.  The resulting correlations observed, $\mathbf{p}^\etal$, where $a$ and $b$ refer only to `conclusive' events, are
\begin{eqnarray}
\label{losscorr}
P^\etal(ab|xy)&=&\etal^2 \;P(ab|xy) \nonumber\\
P^\etal(a\phi |xy)&=&\etal\, P_\t{A}(a|x)\; (1-\etal)  \nonumber\\
P^\etal(\phi b|xy)&=&(1-\etal) \;\etal\, P_\t{B}(b|y) \nonumber\\
P^\etal(\phi\phi |xy)&=& (1-\etal)^2 ,
\end{eqnarray}
where $P_\t{A}$ and $P_\t{B}$ denote the marginal probabilities detected by Alice and Bob in the ideal case (without loss). For simplicity, we take the local efficiencies equal for Alice and Bob and for all measurement settings, but the results can be easily generalized to non-equal local efficiencies. 

In any DIQKD protocol, Alice and Bob construct the key from the outputs of $n_\t{k}$ pairs of measurement settings (typically $n_\t{k}=1$, and the key is generated from the pair ($x^*,y^*$)~\cite{gisin2010,curty2011,meyerscott2013,seshadreesan2016,pitkanen2011}). In \citeref{acin2015b}, a successful eavesdropping attack was constructed for a critical value of the losses equal to $\etac=1/(n_\t{k}+1)$ if  $n_\t{k}<m$, where $m$ is the total number of measurement settings, and $\etac=1/m$ when $n_\t{k}=m$. To implement this attack, Eve needs to be able to control the detection efficiencies on one side, say Alice. Eve has perfect knowledge of the outputs of the $n_\t{k}$ measurements used by Alice for the key, while reproducing the expected correlations~\eqref{losscorr} for $\etal=\etac$. 

For detection efficiencies above this critical value, $\etal>\etac$, Eve can use a combined strategy, in which the previous attack is applied on Alice's side with probability $q_\t{A}$, while with probability $1-q_\t{A}$ Eve does nothing on the measured state and shifts Alice's detection efficiency to one. This attack produces correlations between Alice and Bob of the form~\eqref{losscorr} when $q_\t{A}$ is chosen such that $q_\t{A}+(1-q_\t{A})\etac=\etal$. 

At present, the asymptotic secret-key rate $R$ of any DIQKD protocol in which the key is established by one-way classical communication reconciliation techniques is determined by the best-known lower bound of \citet{friedman2016}, which is valid for most general eavesdropping attacks and reads:
\begin{equation}
R
\quad\ge\quad
\rateH
=
H(\t{A}|\t{E}) - H(x^*|y^*) .
\label{eq.secret.col}
\end{equation}
Here $H(x^*|y^*)$ is the classical conditional Shannon entropy between Alice and Bob outputs when choosing any inputs $(x^*,y^*)$ used for the key and $H(\t{A}|\t{E})$ is the conditional von Neumann entropy between Alice's output and the quantum state in the hands of the eavesdropper, Eve. Crucially, the Bell violation observed by Alice and Bob allows them then to estimate (lower-bound) $H(\t{A}|\t{E})$ without making any assumptions about Eve~\cite{acin2007,pironio2009,masanes2011,pironio2012,vazirani2012,friedman2016}.

\begin{figure}[t!]
\centering
\includegraphics[width=1.0\linewidth]{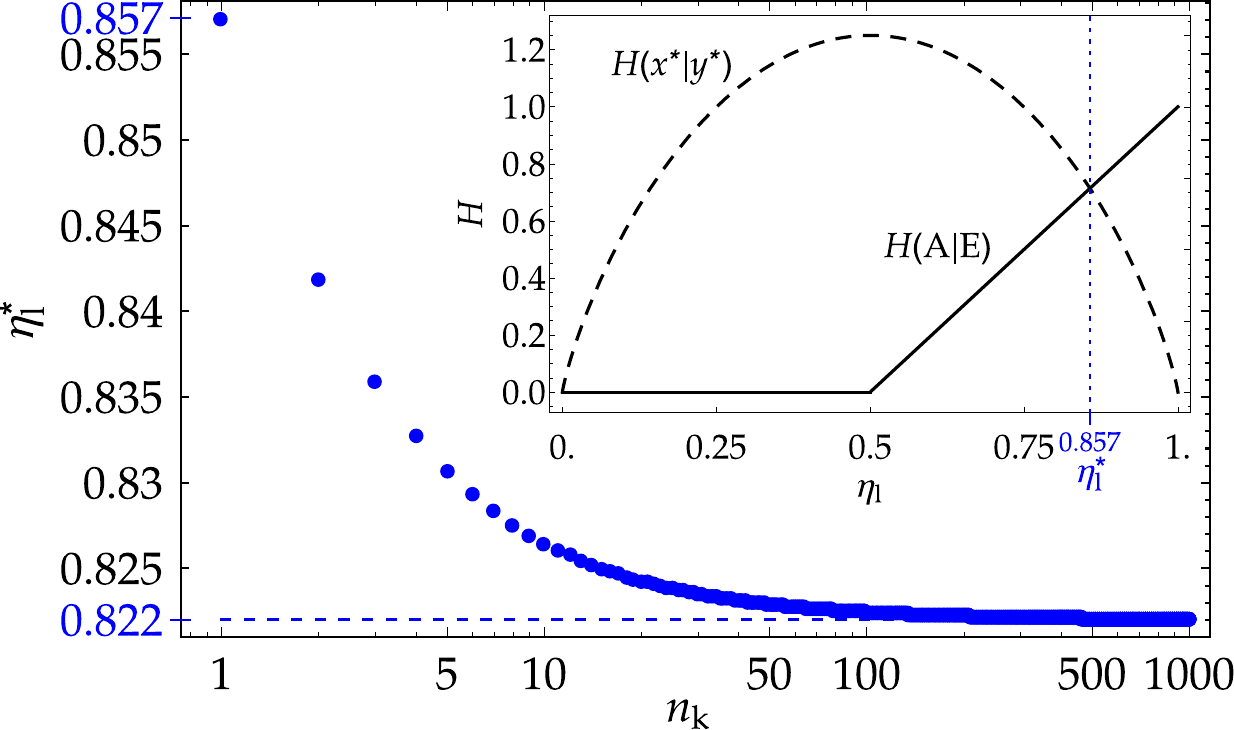}
\caption {%
\textbf{Lower bound on critical local efficiency}, $\etalC$, \textbf{for DIQKD} as a function of the number of measurement settings, $n_\t{k}$, that are used to generate the key. For each $n_\t{k}$ and any $\etal<\etalC$ below the corresponding value (blue dot), there exists a simple attack based on the eavesdropping strategy introduced in \citeref{acin2015b} that prevents any protocol based on two-party correlations \eref{losscorr} from being secure. The most and the least optimistic critical efficiencies for $n_\t{k}\to\infty$ and $n_\t{k}=1$, respectively, are also marked in blue. For $n_\t{k}=1$, the conditional entropies whose difference determines the key rate~\eqref{eq.secret.col} are explicitly shown in the inset.}
\label{fig:simple_attack} 
\end{figure}

However, as \eqnref{eq.secret.col} applies to any attack, we can explicitly evaluate it for the strategy discussed above. Returning to correlations~\eqref{losscorr} that incorporate losses, we compute the conditional entropy $H(x^*|y^*)$. For the sake of simplicity, we perform this calculation for the common case of two-output measurements, while the correlations between Alice and Bob define a perfectly correlated bit in the absence of losses, so that $H(x^*|y^*)=0$ for $\etal=1$. For the above simple attack, we easily see that $H(\t{A}|\t{E})= q_\t{A} \!=\! (\etal - \etac)/(1-\etac)$ for $\etal \geq \etac$, as Eve has then complete knowledge with probability $(1-q_\t{A})$ and complete uncertainty with probability $q_\t{A}$. In contrast, for $\etal\le\eta_c$ the attack of Eve works all of the time, so that then $H(\t{A}|\t{E})=0$. Within the inset of \figref{fig:simple_attack} we plot explicitly both these conditional entropies for $n_k=1$ and $m=2$.

In \figref{fig:simple_attack}, we depict the critical values of the local efficiency, $\etalC$, at which the key rate computed through~\eqref{eq.secret.col} becomes zero as a function of the number of bases, $n_\t{k}$, used to construct the key. Note that based on such an attack, the tolerable local efficiency is forced to be at least $85.7\%$ for any DIQKD scheme with $n_\t{k}=1$, e.g.~the ones of Refs.~\cite{gisin2010,curty2011,meyerscott2013,seshadreesan2016,pitkanen2011} employing one-way communication. Moreover, the above simple attack---with its corresponding critical local efficiencies applying to \emph{any} DIQKD protocol and any Bell inequality which uses the security proof of \citeref{friedman2016}---demonstrates that even in the unrealistic case of users employing an infinite number of bases $n_\t{k}\!\to\!\infty$ (see \figref{fig:simple_attack}) the local efficiencies must necessarily exceed $82.2\%$ for a positive key rate to be possible. %Let us also note that above values are more restrictive than ones stated in Ref.~\cite{gisin2010}, as the latter were obtained by considering a modified version of \eqnref{eq.secret.col}---claimed to allow for some two-way communication between the parties. Unfortunately, as we show in the next section and~\appref{app:Erik}, the security cannot be then generally ensured. 

In view of these results, we expect that high efficiencies will inevitably be needed in DIQKD protocols, and the only solution we foresee is to develop even more efficient photon sources~\cite{Arcari2014,Wein2017,Gustin2017}, better detectors~\cite{marsili_detecting_2013,zhang_advances_2015,miki_stable_2017,moshkova_high-performance_2019} and improve all the couplings within optical implementations to sufficiently decrease losses within the users' laboratories. Nevertheless, we expect Bell experiments with local losses of the order of $90\%$ to be within reach in the near future. In this work, we work under this assumption, which is currently essential for any existing DIQKD implementation.

\subsection{Transmission losses}
\label{sec:trans_losses_heralding}
\begin{figure}[t!]
\centering
\includegraphics[width=1.0\linewidth]{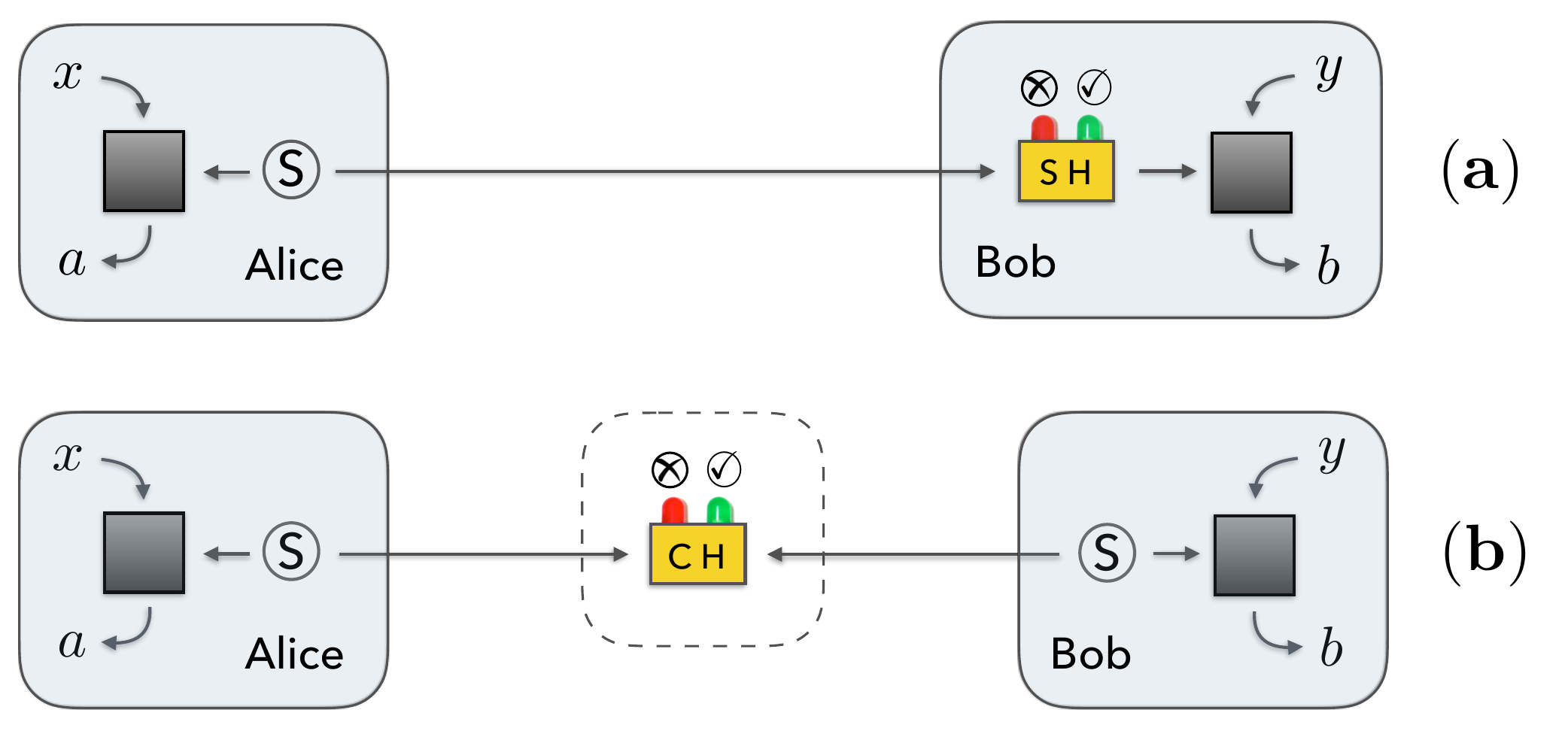}
\caption {%
\textbf{Efficient heralding schemes for DIQKD.}
Alice and Bob are located at isolated labs (shaded regions) from which they 
control information leaks. They locally use sources $\textsf{S}$, to distribute a quantum state between their labs and perform on it randomly sampled measurements labelled $x$ and $y$, producing outcomes $a$ and $b$. The measurement devices are treated as black-boxes that yield a joint probability distribution $\mathbf{p}=\{P(ab|xy)\}_{abxy}$ compatible with the laws of quantum physics. A \emph{heralding scheme} is implemented, such that, given its positive outcome \herald{}, 
the resulting $p(ab|xy\herald)$ shared between Alice and Bob becomes effectively 
independent of the finite transmission efficiency. In the \emph{side-heralding} (SH) scenario, $(\textbf{a})$, this is achieved by one of the users performing a (probabilistic) quantum non-demolition measurement (QND) within their isolated lab that verifies the arrival of the distributed state, without disturbing it. In the \emph{central-heralding} (CH) scheme, $(\textbf{b})$, the heralding is performed by a third party that later publicly announces the successful rounds that should be used during the protocol.}
\label{fig:diqkd} 
\end{figure}

The second type of losses occur while photons propagate outside the labs and are quantified by the \emph{transmission efficiency}, $\etat$, of the channel connecting the users. In principle, they constitute the main hurdle for long-distance DIQKD, as $\etat$ decreases rapidly with distance, e.g.~exponentially when transmitting signals over optical fibres. Moreover, even if fibre technology progresses, the exponential increase of losses with distance will remain, due to unavoidable light absorption and scattering. However, contrary to local losses, transmission losses can be completely overcome by adopting a carefully constructed protocol. A viable route to do so is to record an additional outcome, denoted by \herald{}, indicating in a heralded way that the photons did not get lost~\cite{gisin2010,mattar2016,ZHOU202012}. Then, \herald{} assures that the required quantum state was successfully transmitted
between Alice and Bob and the Bell test can be performed. If the heralding outcome is causally disconnected from the choices of measurement settings $x,y$ by Alice and Bob during each round of the protocol (see \figref{fig:diqkd}), transmission losses become irrelevant with respect to the security of the protocol, affecting only the key rate. In fact, the heralding signal, when causally disconnected from the choice of measurements, can simply be interpreted as a probabilistic preparation of the required state which does not affect a Bell test, nor any protocols based on it.

The heralding process can in principle be implemented with the help of a quantum non-demolition (QND) measurement allowing the number of photons to be measured without disturbing the quantum state~\cite{Grangier1998}. QND photon measurements are, however, challenging, requiring e.g.~unrealistic optical non-linearities. The solution is to replace them with optical linear circuits that achieve the same goal in a probabilistic fashion~\cite{Jacobs2002,Kok2002}. The heralding signal \herald{} is then provided by a particular detection pattern in the linear optics circuit indicating, as for the QND measurement, that the outputs produced by the Bell test are valid.

Within the \emph{side-heralding} (SH) scenario depicted in \figref{fig:diqkd}(\textbf{a}), the circuit is performed by one of the users who records the rounds in which the positive heralding pattern, \herald, has occurred, so that only these are later used for key extraction.
In contrast, in the \emph{central-heralding} (CH) scenario the heralding is performed outside of the users labs, at a central station (resembling the entanglement swapping configuration~\cite{zeilinger1998}), by a third party who then publicly announces which rounds should be considered successful, as illustrated in \figref{fig:diqkd}(\textbf{b}). In either case, the heralding scheme should be causally disconnected from the measurements in the Bell test. This condition is more natural in the CH scheme, within which it is naturally assured by the lack of information leakage from the secure user labs. On the contrary, in the SH configuration it becomes the responsibility of the user holding the heralding device within their lab, who must then, e.g.~ensure that the heralding signal occurs before a random choice of measurement is made\footnote{Ideally, each user possesses an independent source of private randomness~\cite{pironio2010}.}. In any case, the heralding signal should work as the ideal QND measurement and assure that, up to the leading order, transmission losses have no effect on the heralded Bell violation.

The importance of this requirement is best understood by considering existing proposals for photonic DIQKD that do not satisfy it, such as the schemes using a noiseless qubit amplifier~\cite{gisin2010} or entanglement swapping relays~\cite{curty2011,meyerscott2013,seshadreesan2016}. In all these schemes, entanglement between users is distributed using spontaneous parametric down-conversion~\cite{Kwiat1995} (SPDC) sources---a probabilistic process in which multi-photon pair creation also takes place. For the sake of argument, let us consider the
state produced by the SPDC to read:~$\vac\bra{0} + \bar p \ket{\psi_\t{AB}}\bra{\psi_\t{AB}}$;~after, without loss of generality, ignoring its normalisation and the higher-order terms in $\bar p$, i.e.~the spurious contributions arising when more than one photon-pair is created within the process---see \appref{app:states_sources}. For all the schemes,
%of them, 
the state shared by the users after a successful heralding takes a general form (up to irrelevant normalisation):
\begin{equation}
\rho_{\text{AB}|\text{\herald}} = \vac\bra{0} + \lambda\,{\etat}\, \bar{p}\, \ket{\psi_\t{AB}}\bra{\psi_\t{AB}} + \dots,
\label{e:with vac}
\end{equation}
in which the detrimental terms of order $\bar p$ that yield deviations from the target $\ket{\psi_\t{AB}}$ may also be omitted. The parameter $\lambda>0$ above is determined by the particular heralding scheme~\cite{gisin2010,curty2011,meyerscott2013,seshadreesan2016}, while ${\etat}$ is the transmission efficiency dependent on the distance between the users. 

The key point is to notice that the contribution of the maximally entangled state, $\ket{\psi_\t{AB}}$, in \eqnref{e:with vac} occurs at a higher order in $\bar p$ than the vacuum contribution and is influenced differently  by the presence of transmission losses. Thus, for any fixed $\lambda$, the Bell violation strongly depends on ${\etat}$. That is, contrary to when performing an ideal QND measurement, transmission losses not only affect the key rate but also the protocol security. In optical fibres, ${\etat}$ vanishes exponentially with the separation distance $L$ (${\etat} = e^{-L/L_{\t{att}}}$, with typical values of the attenuation length $L_{\text{att}} \approx 20\,\t{km}$), and so the heralded state \eref{e:with vac} approaches the vacuum exponentially with $L$, while rapidly ceasing to produce large enough Bell violations for DIQKD to be possible~\cite{pitkanen2011}. In particular, this implies that in all such protocols there is always a critical distance at which the protocol ceases to be secure.

For the sake of clarity, we provide some simple estimations that make this point more explicit for the scheme based on the qubit amplifier of~\citeref{gisin2010}. As discussed also in \appref{app:state_amplifier}, if one approximates for simplicity $1-\bar p\approx 1$, the state after heralding can be put in the form of \eqnref{e:with vac} with $\lambda=T/(1-T)$, where $T$ is the transmittance of the beam-splitter used in the qubit amplifier for the heralding process~\cite{gisin2010}. Even if quite optimistic, we can take a value of $T=1-10^{-2}$, which gives $\lambda\approx 10^2$. This severely affects the key rate, which is a function of $1-T$, but here we focus on the protocol security. 

For any protocol based on the violation of the CHSH inequality $S_\t{loc}\le2$ (e.g.~the one of \citeref{gisin2010}), the heralded state~\eref{e:with vac} must lead to
\begin{equation}
S\leq \frac{1}{1+\lambda\,{\etat}\, \bar{p}}(2+\lambda\,{\etat}\, \bar{p}\,2\sqrt 2).
\label{eq:CHSH_UB_QA}
\end{equation}
Following \citeref{acin2007}, the inequality~\eqref{eq:CHSH_UB_QA} allows one to lower-bound the term
$H(\t{A}|\t{E})\ge 1 -\chi(S)$ 
in~\eqnref{eq.secret.col}, where $\chi(S)=h[(1+\sqrt{(S/2)^2-1})/2]$ is the binary entropy. 
On the other hand, for key-generation rounds, the conditional entropy $H(x^*|y^*)\!=\!1-h[\bar{\lambda}]$ where $\bar{\lambda}\!=\!\lambda{\etat}\bar{p}/(1+\lambda{\etat}\bar{p})$ is the effective probability of sharing the target state $\ket{\psi_\t{AB}}$, given the state~\eqref{e:with vac}. Putting all these terms together, using the expression for the losses as an exponential function of the distance, and taking an optimistic value of $\bar{p}=10^{-2}$ for SPDC~\cite{giustina2015,lynden2015}, the key rate~\eqref{eq.secret.col} vanishes already for distances of approximately one attenuation length,  $L_{\text{att}} \approx 20\,\t{km}$. 
This critical limit on user separation can be improved by taking even smaller values of $T$ or larger values of $\bar p$, but the problem still remains: for any given values, the weight of the entangled part in the state obtained after successful heralding always decreases exponentially with distance.

The key rates reported in~\citeref{gisin2010} are much higher than those obtained in the above. This arises due to the fact that the authors of \citeref{gisin2010} make additional assumptions on the attacks available to the eavesdropper  (see for example the Supp.~Mat.~of \citeref{gisin2010} and also the discussion in~\citeref{pitkanen2011}). Using these assumptions, they derive a different bound on the key rate as a function of an observed CHSH Bell violation and the rates at which one or both parties observe inconclusive events. The same bound was later used in the protocols of Refs.~\cite{curty2011,meyerscott2013,seshadreesan2016}. Unfortunately, it is unclear whether these assumptions, and corresponding rates, do not imply a loss of generality. In fact, for a slightly different situation in which losses only occur for one of the observers, these assumptions and corresponding bounds can be explicitly proven not to hold: for some value of the losses they predict a strictly positive secret-key rate, while it is possible to derive an explicit eavesdropping attack that breaks the protocol. The details of this attack are shown in \appref{app:Erik}. This analysis implies that the assumptions used in~\citeref{gisin2010}, and later in Refs.~\cite{curty2011,meyerscott2013,seshadreesan2016}, do not hold in full generality and, therefore, it is unclear to what extent the secret-key rates reported in these works are valid.

In what follows, we propose two DIQKD architectures based on single-photon sources~\cite{aharonovich2016} that crucially do not suffer from the above problems. They are designed such that up to the leading order a pure entangled state is shared between the users upon successful heralding -- independent of their separation (or transmission losses). Our protocols thus behave as the ideal QND measurement and allow high key rates to be maintained over large communication distances. One of the schemes relies solely on single photon sources and a CH-based implementation. Since single-photon sources are still an expensive resource compared to widely used SPDC sources, we furthermore consider a SH-based scheme in which both source-types are used in conjunction. In order to maintain generality and a degree of comparison with the SPDC framework~\cite{Kwiat1995}, each single-photon source is modelled to produce a quantum state that, when ignoring normalisation (see also \appref{app:states_sources}), reads $\sigma_\t{SP}=\sum_{n=1}^{\infty}p^{n-1} \proj{n}$ in the photon-number basis, containing an infinite tail of high-order contributions whose probability is parametrised by $p$.

%%%%%%%%%%%%%%%%%%%%%%%%%%%%%%%%%%%%%%%%%%%%%%%%%%%%%%%%%%%%%%%%%%%%%%%%%%%%
\section{DIQKD schemes with single-photon sources}
\label{sec:schemes}
\begin{figure*} [ht!]
\centering
\includegraphics[width=0.85\linewidth]{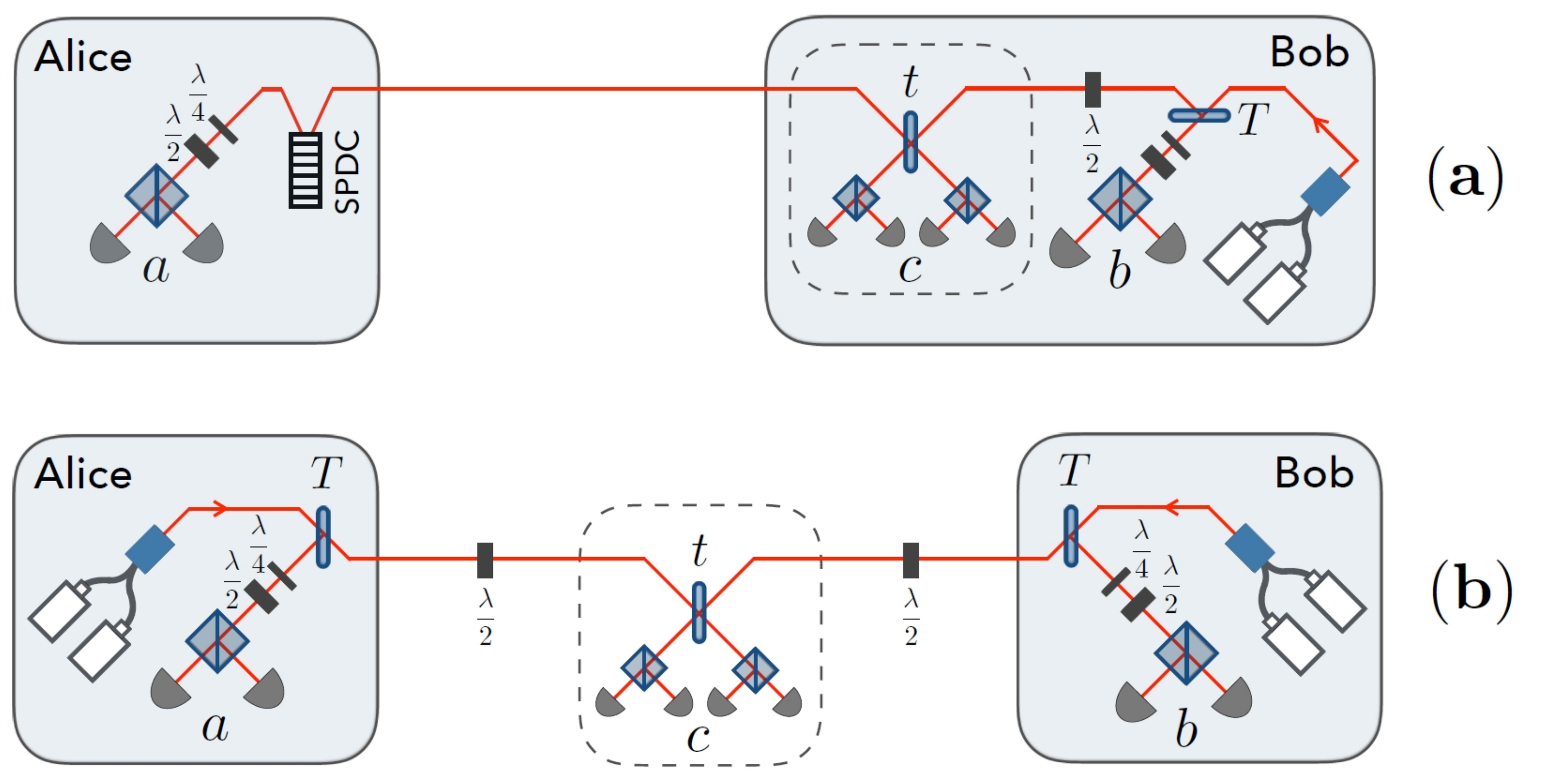}
\footnotesize{\caption {\textbf{DIQKD schemes with single-photon sources.} (\textbf{a}):~\textit{Side-heralding (SH) scheme employing two single-photons sources.} 
The SPDC source is kept close to Alice to avoid transmission losses on her side. Two single photons held by Bob and encoded in orthogonal polarizations impinge a BS 
of transmittance $T\approx1$ located in his lab (shaded region). The reflected mode is then jointly analysed with the system received from Alice by a partial Bell-state measurement (BSM, dashed region) consisting of:~a partial BS of transmittance $t$, polarizing-BSs (split squares) and binary on/off  photodetectors (half-circles).
(\textbf{b}):~\textit{Central-heralding (CH) scheme employing two single-photons sources.} Both Alice and Bob project two single photons encoded in orthogonal polarizations 
onto BSs of transmittance $T\approx0$ situated in their labs. A partial BSM is conducted this time at the central station on the combined polarization components
arriving from Alice and Bob after being passed through HWPs ($\frac{\lambda}{2}$).  In both schemes, the users apply their choice of the measurement settings $x$ and $y$ on the output modes by means of a polarization analyser---a sequence of a quarter-wave plate ($\frac{\lambda}{4}$), HWP, polarizing BS and two binary detectors. } 
\label{fig:scheme}}  
\end{figure*}
The \textit{SH scheme} requires Bob to produce two single photons with orthogonal polarizations $H$ and $V$, while Alice has access to 
entangled photon-pairs produced by an SPDC source. It is inspired by the qubit-amplifier implementation of~\citet{pitkanen2011}, as shown 
in \figref{fig:scheme}(\textbf{a}).  Bob's photons enter a beam-splitter (BS) of transmittance $T$. Then, the reflected light component passes 
through a half-wave-plate (HWP) before being detected in conjunction with Alice's transmitted photons 
via a partial Bell-state measurement (BSM) depicted by the dashed region. The outcome of the BSM, $c$, signifies whether the required
heralding pattern, $c=\herald$, has occurred, corresponding to two detector clicks that represent simultaneous detection of orthogonal polarizations.

Provided that the BS transmittance is kept close to one ($T\approx 1$), \herald{} occurs only when exactly one photon is 
transmitted by the BS while the other photon is reflected, and the single photon-pair term of the
state produced in the SPDC by Alice reaches the BSM. In this manner, the photons distributed to Alice and Bob 
are prepared with orthogonal polarizations, although the information about their concrete polarization is erased 
by the partial BSM. The resulting state shared by Alice and Bob conditioned on $\herald$ corresponds to a 
partially (polarization-) entangled two-qubit state with asymmetry dictated by the BSM transmittance parameter $t$ 
(in an unnormalised form):
\begin{equation}
\label{eq.SHstate}
\rho_{\text{AB}|\herald}^{(\mathrm{SH})} =  \frac{ {\etat}T(1-T)}{8}  \ketbra{\psi^t_\t{AB}}{\psi^t_\t{AB}}  + \bigO{\bar{p}},
\end{equation}
where $\bar{p}$ parametrises the probability to produce multiple pairs in the SPDC process of Alice (see \appref{app:states_CHSH}). 
The target state $\ket{\psi^t_\t{AB}} = \ket{\psi^-_\t{AB}} + t\ket{\phi^-_\t{AB}}$ in \eqnref{eq.SHstate} is a superposition of Bell states given, 
in second quantization, by  $\ket{\psi^-_\t{AB}} = \frac{1}{\sqrt{2}}(a_H^\dagger b_V^\dagger -  a_V^\dagger b_H^\dagger ) \vac$ and $\ket{\phi^-_\t{AB}} = \frac{1}{\sqrt{2}}(a_H^\dagger b_H^\dagger -  a_V^\dagger b_V^\dagger ) \vac$.

\begin{table*}[t!]
\centering
\ra{1}
\begin{tabular}{ r c c }  
\toprule[1pt]
\multirow{ 2}{*}{\textbf{DIQKD Scheme:}} & \parbox[t]{3cm}{\centering{Side-heralding \\ (SH)}} & \parbox[t]{3cm}{\centering{Central-heralding \\ (CH)}}\\
\rule{0pt}{4ex}
\textit{Critical local efficiency $\etalC$ (diqkd)}   &\  94.9\%  & \ 94.3\% \\
\textit{Critical local efficiency $\etalC$ (nonloc.)}   &  74.3\%  &  69.2\% \\
\textit{Noise robustness (nonloc.)}   &  31.2\%  &  35.7\% \\
\textit{Secret key per heralded round (bit fraction $\le1$)}   &  0.82  &  0.95\ \\
\bottomrule[1pt]
\end{tabular}
\caption{\textbf{Performance of DIQKD schemes.} 
Critical local efficiencies, $\etalC$, only above which the secret key can be distributed in a fully device-independent fashion, compared with ones above which the shared correlations exhibit nonlocality. For perfect local efficiencies ($\etal=1$), robustness to mixing the joint probability distribution with a maximally uncorrelated one is listed, as well as the bit fraction of the secret key generated per successfully heralded round, equal to one in the ideal case. The probability of producing a single photon or an SPDC photon-pair is assumed as $ p=\bar p= 10^{-4}$ for each source. 
}
\label{tab.1}
\end{table*}

The \textit{CH scheme} depicted in \figref{fig:scheme}(\textbf{b}) requires both Alice and Bob to produce two single photons with orthogonal polarizations $H$ and $V$, inspired by the entanglement distribution scheme of~\citet{lasota2014}. The photons produced on each side impinge separate BSs of low transmittance ($T\approx0$) and, thus, reach the central station with low probability. The heralding is again provided by a partial BSM, performed now by a third party, after passing both incoming beams through separate HWPs. The signal \herald{} is observed only when each party transmits exactly one single photon and in such a case the reflected photons kept by Alice and Bob are again in a partially (polarization-) entangled state with  asymmetry determined by the transmittance $t$ of 
the partial BSM performed at the central station (see \appref{app:states_CHSH}):
\begin{equation}
\label{eq.CHstate}
\rho_{\text{AB}|\herald}^{(\mathrm{CH})} = \frac{{\etat}T^2(1-T)^2}{4} \ketbra{\psi^t_\t{AB}}{\psi^t_\t{AB}}  + \bigO{p}.
\end{equation}

Unlike previous proposals, see \eqnref{e:with vac}, in the above two schemes the vacuum terms do not emerge after heralding. Moreover, the unnormalised states \eqref{eq.SHstate} and \eqref{eq.CHstate}, to first significant order, are pure and proportional to the transmission efficiency ${\etat}$. This guarantees that, after normalisation, the states are independent of ${\etat}$ (to first order). This, and the use of single-photon sources instead of SPDC, are the crucial ingredients that allow us to achieve significantly higher secret key rates at larger distances than previous proposals.

The second advantage of our schemes is that, by adjusting the transmittance $t$ of the partial BSM, the entanglement of the target state $\ket{\psi^t_\t{AB}}$ can be continuously tuned between the maximally entangled ($t=0$) and product ($t=1$) extremes~\cite{eberhard1993}. This can then be used, in particular, to improve the local efficiencies, $\etal$, required to meet the security requirements.

We notice that it is possible to reduce the number of single photon sources in our schemes by using a single source emitting a temporal stream of  photons. This would require, however, the stream to be de-multiplexed either by active optics (which would add extra noise) or probabilistically by passive elements (which would decrease the final heralding rate).

%%%%%%%%%%%%%%%%%%%%%%%%%%%%%%%%%%%%%%%%%%%%%%%%%%%%%%%%%%%%%%%%%%%%%%%%%%%%
\section{Computing key rates in heralded schemes}
\label{sec:key_rates}
In standard DIQKD protocols, Alice and Bob measure their particles. A subset of these measurements is publicly announced so that the users can count how many times different outcomes ($a$,~$b$) are obtained for the different combinations of inputs ($x$,~$y$). From this information, they compute the amount of achievable secret key and, if positive, distil it by means of classical post-processing~\cite{Scarani2009} from the remainder of data being shared, specifically, from particular pre-designed measurement settings ($x^*$,~$y^*$)~\cite{acin2007,pironio2009,masanes2011}. As mentioned, in the asymptotic limit of infinitely many rounds, the attainable key rate is given by \eqnref{eq.secret.col}, which constitutes the best known lower bound, that, crucially, is valid for the most general eavesdropping attacks~\cite{friedman2016}.

Calculating exactly $H(A|E)$ in \eqnref{eq.secret.col}, for a given observed correlation $\mathbf{p}$ or Bell inequality violation, optimising over attacks of Eve, turns out to be extremely hard. The problem has only been solved for the CHSH inequality~\cite{acin2007}, the simplest of all Bell inequalities. Here, we use a lower bound on $H(A|E)$, which in turn provides a lower bound on the key rate, computable for any type of correlations. It is obtained by replacing the von Neumann entropy in \eqnref{eq.secret.col} by the min-entropy~\cite{masanes2011}. This quantity is then directly connected to the guessing probability, $G_{\mathbf{p}}(x^*)$, for Eve to correctly guess Alice's output when she performs the measurement $x^*$. It can be computed for any Bell correlations exhibited by $\mathbf{p}$ by means of semi-definite programming, as explained in \appref{app:pguess}. The resulting bound on the key rate~\eqref{eq.secret.col}, which has already appeared in previous security proofs~\cite{masanes2011,pironio2012}, reads 
\begin{equation}
\label{eq.secret}
R\;\ge\;\rateH\;\ge\;
\rate=-\log_2 G_{\mathbf{p}}(x^*)- H(x^*|y^*).
\end{equation}
In an ideal scenario with two outcomes, there are no errors between Alice and Bob, $H(x^*|y^*)=0$, and Eve has no information about Alice's outputs, $G_{\mathbf{p}}(x^*)=1/2$, so that $R=\rate=1$. Because of its ease of computation, $\rate$ is the quantity used here to estimate attainable key rates of the implementations proposed.

When considering protocols that incorporate a heralding stage depending on an outcome $c$, with rounds occurring at a repetition rate $\nu_{\t{rep}}$, we quantify the effective key rate of secret bits certified per time-unit as:
\begin{equation}
K\ =\ \nu_{\t{rep}}\  P(c\!=\!\herald)\   \rate,
\label{keyDIQKD}
\end{equation}
where $P(c\!=\!\herald)$ is the probability of successful heralding in each round.

%%%%%%%%%%%%%%%%%%%%%%%%%%%%%%%%%%%%%%%%%%%%%%%%%%%%%%%%%%%%%%%%%%%%%%%%%%%%
\section{Performance of the SH and CH schemes}
\label{sec:results}
As a result, we may quantify the optimal DIQKD-performance of the CH and SH schemes depicted in \figref{fig:scheme}, by conducting an unconstrained nonlinear maximisation of $K$ in \eqnref{keyDIQKD} over all adjustable parameters. In particular, for each of the schemes, we optimise  over the source parameters $p$ and $\bar{p}$, transmittance values $T$ and $t$, as well as polarization angles specifying the user measurements. Still, we ensure $p,\bar{p}\approx0$ and $T\approx1$ (or $T\approx0$) in case of the SH (or CH) scheme, so that the distributed quantum states can be truncated at a finite order in $p$, $\bar{p}$ and $1-T$ (or $T$). Nonetheless, in order to maintain security we bypass such a truncation by giving full control to the eavesdropper over the higher-order terms that are neglected. Moreover, the critical noise parameters can then also be determined by similar optimisation procedures---conducted while increasing the noise until the key rate \eref{keyDIQKD} cannot be made strictly positive. Explicit details about these optimization steps are given in Apps.~\ref{app:pguess}, \ref{app:higher_terms} and \ref{app:robust}.

We summarise the performance of the SH and CH schemes in \tabref{tab.1}. Although the SH scheme is simpler, requiring only two single-photon sources, its performance is worse than the CH scheme for all figures of merit considered. From here onwards, we thus focus on the CH scheme, which defines the ultimate experimental requirements for DIQKD to be possible within our approach. In what follows, we show that this scheme offers reasonable levels of robustness against all relevant noise parameters.

The resistance to noise is estimated using a simple noise model, in which the ideal correlations are mixed with white-noise correlations with weight $1-v$ and $v$, and perfect local efficiency ($\etal=1$) is assumed. The CH scheme yields nonlocal correlations up to $v=35.7\%$ level of mixing. Concerning imperfection of the single-photon sources, for a realistic value of multi-photon generation of $p=10^{-4}$ (c.f.~\cite{ding2016}) the CH scheme generates up to $0.95$ secret bits per (successfully) heralded round---achieving close to the ultimate limit of $1$ secret bit, applicable in a perfectly noiseless scenario~\cite{acin2012}. The critical local efficiencies, $\etalC$ for the nonlocality to be observed are very close to the ultimate bound of \citet{eberhard1993}, $\etal\ge66.(6)\%$, which can be approached due to the ability to prepare pure partially entangled two-qubit states within both the SH and CH schemes.

Most importantly, employing the CH scheme for DIQKD, our work predicts that positive key rates can be generated independently of the separation between Alice and Bob, as long as the effective local efficiency, $\etal$, for each of the user labs is higher than $94.3\%$. Assuming $\etal$ to be the product of the efficiencies of:~photon extraction from each single-photon source employed ($\eta_\t{ls}$), transmission between the sources and detectors involved ($\eta_\t{lt}$), and detection ($\eta_\t{ld}$);~fully secure DIQKD is possible as long as $\etal=\eta_\t{ls}\eta_\t{lt}\eta_\t{ld}\ge0.943$ can be attained by each user.

\begin{figure}[!t]
\includegraphics[width=1\linewidth]{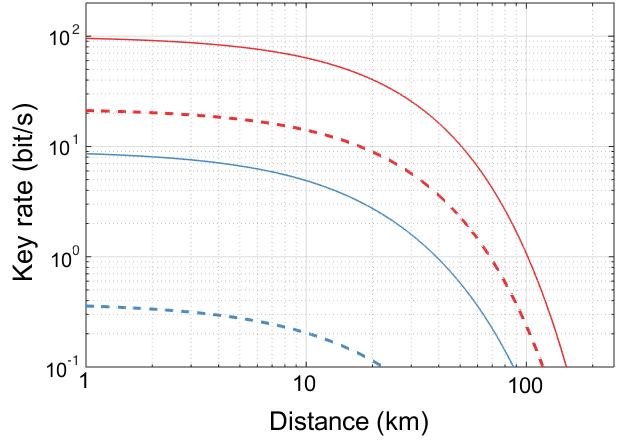}
\caption{%
\textbf{DIQKD key rates attained with $95\%$} (blue) \textbf{and $96\%$} (red) \textbf{local efficiencies.} In each case, the solid (dashed) curve represents the key rate in bits per second attained by the CH (SH) scheme. Each key rate is optimised over all adjustable physical parameters, yet in the case of the single-photon sources impurity parameter, $p$, its lowest possible value is always favoured. Here, we fix $p=10^{-4}$, and consider the repetition rate of photon extraction for each source to be $100$ MHz~\cite{ding2016}.}
\label{fig.secretrate}
\end{figure}

Taking, for instance, $\etal=95\%$ (see also \figref{fig.secretrate}), the secret key can be securely distributed over large distances while completely avoiding the transmission losses. In particular, assuming in \figref{fig.secretrate} for the CH scheme:~realistic ${\etat} = \e^{-L/L_{\t{att}}}$ with $L_{\t{att}}= 22\,\t{km}$~\cite{gisin2010}, the lab beam-splitters to exhibit transmission $T\approx10^{-3}$, and each of the sources employed to be producing photons at $100$~MHz rate with $p=10^{-4}$~\cite{ding2016};~a key rate of $1\,\t{bit/s}$ can be attained over approximately $50\,\t{km}$. In \figref{fig.secretrate}, we consider also the SH scheme, for which we then assume the SPDC source to produce entangled photons also at $100$~MHz rate with $\bar{p}\approx10^{-4}$---both theoretically within current technological reach~\cite{zhang2007,giustina2015,lynden2015}---but also ensure $T\approx1-10^{-3}$ to make the comparison fair. 

However, let us note that the corresponding values of key-rates are primarily dictated by the factor $K\propto\nu_\t{rep}\,P(c\!=\!\herald)$ appearing in \eqnref{keyDIQKD} with the successful-heralding probability effectively equal to $T^2$ and $\bar{p}\,(1-T)$ at $L=0$ for the CH and the SH scheme, respectively. In particular, as for the single-photon sources we take $p>0$ to account solely for multi-photon events, its impact on the key rates is negligible. Although in our analysis we were motivated to use the least number of single-photon sources, under such an assumption the limiting dependence of key rates on how well $T\approx0$ (or $T\approx1$) could, in principle, be avoided by creating more photons and performing within each lab extra (local) pre-heralding~\cite{Jacobs2002,Kok2002}, which must importantly assure a polarisation-entangled photon pair to be distributed. When employing multiple SPDC sources such an approach may seem to be even less efficient~\cite{sliwa_conditional_2003,barz_heralded_2010,wagenknecht_experimental_2010}, however, rapid development of solid-state emitters capable of producing on-demand entangled photons could provide a breakthrough~\cite{salter_entangled-light-emitting_2010,stevenson_indistinguishable_2012,trotta_highly_2014}.

%%%%%%%%%%%%%%%%%%%%%%%%%%%%%%%%%%%%%%%%%%%%%%%%%%%%%%%%%%%%%%%%%%%%%%%%%%%%
\section{Conclusions and outlook}
\label{sec:conclusions}
Two proposals for photonic implementations of DIQKD schemes have been given here. They make use of side- or central-heralding, and utilise two or four single-photon sources, respectively. They are capable of maintaining security despite arbitrary transmission losses, and distribute keys over large distances given sufficiently high local efficiencies. 

The analysis proves the proposed photonic architectures to be almost optimal from the implementation perspective, as they allow to nearly perfectly compensate for the impact of finite transmission, so that devices can operate independently of the distance separating the users. In contrast, as shown to be generally demanded within the context of DIQKD, being a feature of current state-of-the-art security proofs~\cite{friedman2016}, the requirements on local efficiencies for the protocols remain to be stringent. Hence, an important question that remains open is whether these demanding requirements can be improved by developing more elaborate proofs that, in particular, allow the users to perform two-way communication during the protocol rather than only one-way error correction that is typically assumed~\cite{acin2007,pironio2009,masanes2011,pironio2012,vazirani2012,friedman2016}. Unfortunately, recent progress in this direction has indicated that not much room for improvement may be available in this respect without jeopardising the full security~\cite{thinh2015,tan_advantage_2020}.

Another important future direction is to improve the key-rate analysis presented here, while accounting in more detail for the limitations of particular photonic components being employed~\cite{pir2019advances}. On the one hand, an explicit study of the impact of detector dark counts would be valuable, even though our noise-robustness analysis suggests these not to play a major role (see the values presented in \tabref{tab.1}). On the other hand, the protocol repetition-rates have been assumed to be primarily dictated by the capabilities of the photon sources employed~\cite{ding2016,zhang2007}, while ignoring, e.g.~the finite dead-time of the binary detectors.

Nonetheless, while the requirements on local efficiencies for the proposed protocols are currently challenging, based on the rapid technological improvement and anticipated capabilities of single-photon sources~\cite{Arcari2014,Wein2017,Gustin2017} and detectors~\cite{marsili_detecting_2013,zhang_advances_2015,miki_stable_2017,moshkova_high-performance_2019}, we hope that the demands on fully secure DIQKD implementations presented already here will be fulfilled in the future.

\paragraph*{Note Added.}
After making this work publically available online at \href{https://arxiv.org/abs/1803.07089}{arXiv:1803.07089 [quant-ph]}, an alike study of CH and SH schemes for DIQKD has been released~\cite{zapatero_long-distance_2019}, which by focusing on the Bell violation of the CHSH inequality arrives at slightly higher requirements for local efficiencies but goes beyond the asymptotic key-rate analysis---see also a very recent work \cite{murta_towards_2019} that develops finite-key analysis for DIQKD. Moreover, the model of the CH scheme has been developed and explicitly verified against an experimental implementation within the scenario in which Alice and Bob possess an SPDC source of entangled photons each (rather than two single-photon sources)~\cite{tsujimoto_heralded_2019}. Although the theoretical predictions have been demonstrated to accurately reproduce the observed correlations, these currently do not exhibit Bell violations strong enough for DIQKD, as the corresponding local efficiencies do not yet reach the stringent regime of $\etal\gtrsim95\%$ indicated by our work.

%%%%%%%%%%%%%%%%%%%%%%%%%%%%%%%%%%%%%%%%%%%%%%%%%%%%%%%%%%%%%%%%%%%%%%%%%%%%
\begin{acknowledgments}
We thank Rotem Arnom-Friedman, Miko\l{}aj Lasota, Stefano Pironio and Nicolas Sangouard for helpful discussions. This work was supported by the ERC CoG QITBOX and AdG CERQUTE, Spanish MINECO (Severo Ochoa SEV-2015-0522), Fundacio Cellex and Mir-Puig, the AXA Chair in Quantum Information Science, the Generalitat de Catalunya (SGR1381 and CERCA Program), the Royal Society (URF UHQT), the EU Quantum Flagship projects QRANGE and CiviQ, as well as by the Foundation for Polish Science under the ``Quantum Optical Technologies'' project carried out within the International Research Agendas programme co-financed by the European Union under the European Regional Development Fund.
\end{acknowledgments}

%%%%%%%%%%%%%%%%%%%%%%%%%%%%%%%%%%%%%%%%%%%%%%%%%%%%%%%%%%%%%%%%%%%%%%%
% Appendices
%%%%%%%%%%%%%%%%%%%%%%%%%%%%%%%%%%%%%%%%%%%%%%%%%%%%%%%%%%%%%%%%%%%%%%%
%%%%%%%%%%%%%%%%%%%%%%%%%%%%%%%%%%%%%%%%%%%%%%%%%%%%%%%%%%%%%%%%%%%%%%%
%\clearpage
%\noindent
%\onecolumngrid
%{\centering {\Large \textbf{Supplementary Information for}}\\
%\vspace{2mm}
%{\large \FirstAuthor~et al.~``\textit{\MyTitle}'' \\}}
%\vspace{5mm}
%\twocolumngrid
\appendix
%%%%%%%%%%%%%%%%%%%%%%%%%%%%%%%%%%%%%%%%%%%%%%%%%%%%%%%%%%%%%%%%%%%%%%%

%%%%%%%%%%%%%%%%%%%%%%%%%%%%%%%%%%%%%%%%%%%%%%%%%%%%%%%%%%%%%%%%%%%
\section{States produced by the SPDC and single-photon sources} 
\label{app:states_sources}

The process of spontaneous parametric down-conversion~\cite{Kwiat1995} (SPDC) producing two-mode polarisation entangled photons is described by the Hamiltonian $\hat H=\i\kappa(a_{H}^{\dagger}b_{V}^{\dagger}-a_{V}^{\dagger}b_{H}^{\dagger})+h.c.$,  where $a_{H}^{\dagger}$, $a_{V}^{\dagger}$, $b_{H}^{\dagger}$ and $b_{V}^{\dagger}$ are the bosonic creation operators of the two spatial modes $a$ and $b$, with $H$ and $V$ denoting their orthogonal polarizations. Rewriting $\hat{H}$ with help of the $\t{su(1,1)}$ algebra  generators, i.e.~ones that obey $\left[L_{-},L_{+}\right]=2L_{0}$ and $\left[L_{0},L_{\pm}\right]=\pm L_{\pm}$, it is straightforward to verify that the state produced via the SPDC reads~\cite{kok2000}:
\begin{align}
\ket{\Psi_\t{SPDC}}  
& =
\e^{-\i \hat Ht}\vac = \e^{\tau\left(L_{+}-L_{-}\right)}\vac \\
& =\left(1-\tanh^{2}\tau\right)\e^{\tanh\tau L_{+}}\vac,
\end{align}
where $L_+=L_-^\dagger=a_{H}^{\dagger}b_{V}^{\dagger}-a_{V}^{\dagger}b_{H}^{\dagger}$, $\vac$ denotes the vacuum state of all modes, while $\tau=\kappa\,t>0$ can be assumed to be real.

Moreover, as throughout this work we consider photonic schemes based on (binary, on/off) photodetection, the state $\Psi_\t{SPDC}$ should be interpreted as an incoherent mixture of different photon-number states due to lack of a global phase reference. Hence, defining $q=\tanh^2\tau$ as the effective parameter of the SPDC process, one arrives at the expression:
\begin{align}
\varrho_\t{SPDC}  
& = (1-q)^2 \sum_{n=0}^{\infty}\left(n+1\right)q^{n}\proj{\Psi_{n}},
\label{eq.spdc_full}
\end{align}
where $\ket{\Psi_n} = \frac{1}{n!\sqrt{n+1}}L_+^n\vac$ is the pure state obtained when $n$ photon-pair excitations occur during the down-conversion. 

Nonetheless, for simplicity and the purpose of our work, we redefine the state \eref{eq.spdc_full} in an unnormalised fashion as,
\begin{align}
\rho_\t{SPDC}  
&=
\sum_{n=0}^{\infty}\frac{n+1}{2^n}\bar{p}^n \proj{\Psi_n} 
\label{eq.spdc}\\
& = \proj{0} + \bar{p} \ketbra{\Psi_1}{\Psi_1}+\bigO{\bar{p}^2},
\label{eq.spdc_exp}
\end{align}
such that $\t{Tr}[\rho_\t{SPDC}]=1/(1-2\bar{p})^2$, and so that the parameter $\bar{p}=2q$ can now be directly associated with the contribution of the desired singlet: 
\begin{align}
\ket{\Psi_1} & =
\frac{1}{\sqrt 2}\left(\ket{1_H}_a\ket{1_V}_b-\ket{1_V}_a\ket{1_H}_b\right) \\
& = 
\frac{1}{\sqrt 2}\left(\ket{HV}-\ket{VH}\right).
\label{eq.singlet}
\end{align}

Experimentally, the parameter $\bar{p}$ is kept small (below $10^{-2}$) and may be adjusted with squeezing techniques~\cite{giustina2015,lynden2015}. Although large values of $\bar{p}$ increase the production rate of the target maximally entangled two-photon states, $\ket{\Psi_1}$, they also increase the relative contribution of spurious higher-order terms, $\ket{\Psi_{n>1}}$, to the SPDC process.

On the other hand, as stated in the main text, whenever the single-photon (SP) sources~\cite{aharonovich2016} are used, we represent states they produce in an analogous unnormalised, $\t{Tr}[\sigma_\t{SP}]=1/(1-p)$, manner as:
\begin{align}
\sigma_\t{SP}  
&=
\sum_{n=1}^{\infty}p^{n-1} \proj{n}
\label{eq.sp}\\
& = \proj{1} + p \proj{2}+\bigO{p^2},
\label{eq.sp_exp}
\end{align}
where the desired single-photon is then produced at the \emph{zeroth} order in $p$ ($\approx10^{-4}$ in current experiments~\cite{ding2016})---in contrast to the SPDC process \eref{eq.spdc} in which the target photon-pair \eref{eq.singlet} occurs at the \emph{first} order in $\bar{p}$ in \eqnref{eq.spdc_exp}. 

Finally, let us emphasise that throughout this work we perform calculations for all the schemes beyond their expected ideal working-order in $\bar{p}$ and $p$, i.e.~by performing truncations of \eref{eq.spdc_exp} and \eref{eq.sp_exp} at higher orders. Still, it is crucial to mention that, when we compute the results (key rates and figures of merit presented in \tabref{tab.1} of the main text), we nevertheless bypass such a truncation by assuming that higher-order terms (those which were dropped) are controlled by the eavesdropper to her own benefit. We give the details of this technique in \appref{app:higher_terms} below. 

%%%%%%%%%%%%%%%%%%%%%%%%%%%%%%%%%%%%%%%%%%%%%%%%%%%%%%%%%%%%%%%%%%%
\section{Heralded state produced by the qubit amplifier of \citet{gisin2010}} 
\label{app:state_amplifier}

The original scheme of \citeref{gisin2010} is of the SH type (see \figref{fig:diqkd}(\textbf{a}) of the main text) and 
consists of an SPDC source held by Alice and two single-photon sources (emitting photons in $H$ and $V$ polarisation modes) 
held by Bob. For the sake of the argument, let us assume that all the sources do not produce multiple pairs, which is only beneficial for the scheme. 

The initial composite state of Alice and Bob before communication and amplification reads~\cite{gisin2010}: 
\begin{equation}
\left[(1-\bar{p})\ketbra{0}{0}+\bar{p}\ketbra{\Psi_1}{\Psi_1}\right] \otimes \ketbra{1_H}{1_H}\otimes \ketbra{1_V}{1_V}.
\end{equation} 
Bob's photons enter a beam-splitter of transmittance $T$, so that the reflected mode can then be combined with the mode received from Alice within an implementation of the Bell-state measurement (BSM). As a consequence, the final unnormalized state that is shared by Alice and Bob, conditioned on the (heralding) success of the BSM performed by Bob, reads:
\begin{equation}
(1-\bar{p})(1-T)^2\ketbra{0}{0} \ +\  \bar{p}\etat T(1-T)\ketbra{\Psi_1}{\Psi_1}\ +...,
\label{eq.app1}
\end{equation}
where we have already ignored all irrelevant terms that do not yield any correlations apart from the vacuum---which occurs with probability proportional to $(1-T)^2$, since both of Bob's photons are reflected and detected. The second term in \eqnref{eq.app1} corresponds to the case when Alice produces the singlet \eref{eq.singlet}, which is transmitted with probability $\etat$, and only one of Bob's photons is reflected. One can see that \eqnref{eq.app1} is of the form of \eqnref{e:with vac} in the main text with the effective $\lambda=T/(1-T)$.

Such a feature will always emerge as long as the singlet (target) state is proportional to $\etat$, while the vacuum component remains unaffected by the finite transmission efficiency. In particular, it naturally generalises to scenarios based on `entanglement-swapping' or `teleportation'~\cite{curty2011,meyerscott2013,seshadreesan2016} and hence, as explained in the main text, constitutes the main limitation of all these schemes. The only exception is the `quantum-relay'-based scheme proposed in~\citeref{curty2011} that, however, due to SPDC sources being employed yields a conditional state still containing undesired terms in apart from the singlet contribution in \eqnref{eq.SHstate}.

%%%%%%%%%%%%%%%%%%%%%%%%%%%%%%%%%%%%%%%%%%%%%%%%%%%%%%%%%%%%%%%%%%%
\section{Secret-key rate under losses} 
\label{app:Erik}

In this appendix, we present a rather natural scenario in which the bound on the key rate derived by \citet{gisin2010} can be proven not to hold. In the supplemental material of \citeref{gisin2010}, the situation is studied in which Alice and Bob implement lossy measurements on an entangled state. The goal is to establish an upper bound on the information that an eavesdropper can possess about the outcomes used for generation of the secret key, given that the non-detected events have already been discarded. 

A bound based only upon the statistics of the conclusive events is not possible, as it would open the detection loophole. In Ref.~\cite{gisin2010} a method is given for bounding Eve's knowledge about the conclusive correlations, based upon the full (lossy) correlations. The main result, see Eq. (10) in their work, is the following bound on the mutual information,
$I(\t{A}:\t{E})=H(\t{A})-H(\t{A}|\t{E})$, between Alice and Eve:
\begin{equation}
I(\t{A:E})\;\leq\; I_\t{E}(S_\t{cc},\mu)=(1-\mu)\,\chi\!\left(\frac{S_\t{cc}-4\mu}{1-\mu}\right)+\mu.
\label{eq:GPSbound}
\end{equation}
Here, $\mu$ is a parameter defined by the ratio of the rates of conclusive-conclusive events, $\mu_\t{cc}$, and conclusive-inconclusive events, $\mu_\t{ci}$ and $\mu_\t{ic}$, from Alice's and Bob's perspective, respectively, and reads
\begin{equation}
\mu=\frac{\mu_\t{ci}+\mu_\t{ic}}{\mu_\t{cc}}.
\end{equation}
The parameter $S_\t{cc}$, on the other hand, denotes the value of the CHSH inequality when computed \emph{only} from the conclusive events. Finally, the function 
\begin{equation}
\chi(S)=h\!\left(\frac{1+\sqrt{(S/2)^2-1}}{2}\right)
\end{equation}
with $h(x)=-x\log_2 x-(1-x)\log_2(1-x)$ has already been employed in the main text, see below \eqnref{eq:CHSH_UB_QA}, and follows from \citeref{acin2007}. For what follows, the property to remember is that $\chi(S)<1$ if $S>2$. We also emphasise that the bound~\eqref{eq:GPSbound} depends on the lossy correlations: while the Bell parameter used in $I_\t{E}$ in \eqnref{eq:GPSbound} is estimated only from the conclusive events, $I_\t{E}$ depends also on the rates of conclusive and inconclusive events via the parameter $\mu$.

Let's apply the bound~\eqref{eq:GPSbound} to a situation in which losses only appear on Alice's side. The corresponding correlations, $\mathbf{p}^\etal$, between Alice and Bob then read analogously to \eqnref{losscorr} of the main text:
\begin{eqnarray}
P^{\etal}(ab|xy)&=&\etal P(ab|xy) \nonumber\\
P^{\etal}(\phi b|xy)&=&(1-\etal) P_\t{B}(b|y) ,
\label{1losscorr}
\end{eqnarray}
where again $a$ and $b$ refer only to conclusive `detection' events, while $P_\t{A}$ and $P_\t{B}$ denote the marginal probabilities detected by Alice and Bob (in the ideal lossless case). We consider the standard situation in which Alice and Bob implement the optimal measurements to violate the CHSH inequality, given a singlet is shared, while the key is generated from one of these measurements, say $x=0$. Therefore, Eve's goal is to guess the output of this measurement on Alice's side.

The correlations~\eqref{1losscorr} can be conveniently arranged in a table as follows
  \begin{equation}
\mathbf{p}^\etal=
\left[
      \begin{array}{cc | cc}
          \etal s & \etal t 
        & \etal s & \etal t \\
          \etal t & \etal s
        & \etal t & \etal s \\ 
        \frac{1 - \etal}{2} & \frac{1 - \etal}{2} &
        \frac{1 - \etal}{2} & \frac{1 - \etal}{2} \\
        \hline
          \etal s & \etal t 
        & \etal t & \etal s \\
          \etal t & \etal s
        & \etal s & \etal t \\ 
        \frac{1 - \etal}{2} & \frac{1 - \etal}{2} &
        \frac{1 - \etal}{2} & \frac{1 - \etal}{2}
        \end{array}
    \right] \,,
 \label{losstsirelson}
  \end{equation}
where within each of the four blocks the columns are labelled by the two possible conclusive outputs of Bob, and the rows by the three outputs of Alice that include the non-detected outcome. The four blocks above correspond then to the four combinations of measurement settings $x,y=\{0,1\}$, where $s=(1+\cos(\pi/4))/4$ and $t=(1-\cos(\pi/4))/4$.  Whenever the local efficiency is unity, $\etal=1$, the correlations~\eqref{losstsirelson} violate maximally the CHSH inequality and are referred to as the `Tsirelson correlations'.
 
It can be verified that the correlations~\eqref{losstsirelson} are local whenever $\etal\leq 1/\sqrt 2$. On the other hand, when using them to evaluate the bound~\eqref{eq:GPSbound} for any of Alice's measurement, say $x=0$, one obtains $I_\t{E}(S_\t{cc},\mu)<1$ whenever $\etal> 1/\sqrt 2$. The result is intuitively satisfactory:~if the initial correlations are non-local, there is some uncertainty left for Eve about Alice's outcome after discarding the non-conclusive events. Unfortunately, this conclusion, and therefore the bound~\eqref{eq:GPSbound} used to derive it, is not universally valid, as proven by the following attack of Eve, whereby she has perfect knowledge of Alice's outcome for some values of $\etal$ larger than $1/ \sqrt{2}$.
 
Eve prepares a mixture of the following three distributions:
\begin{equation}
  \mathbf{p}_{1} = \left[
    \begin{array}{cc|cc}
      s & t & s & t \\
      t & s & t & s \\
      0 & 0 & 0 & 0 \\ \hline
      p s & p t & p t & p s \\
      p t & p s & p s & p t \\
      \tfrac{1}{2} (1 - p) & \tfrac{1}{2} (1 - p)
      & \tfrac{1}{2} (1 - p) & \tfrac{1}{2} (1 - p) 
    \end{array}
  \right]
\end{equation}
with $p = \sqrt{2} - 1 \approx 0.41$, which is local and, hence, can be further decomposed in terms of deterministic strategies;~and
\begin{equation}
  \mathbf{p}_{2} = \left[
    \begin{array}{cc|cc}
      s & t & s & t \\
      0 & 0 & 0 & 0 \\
      t & s & t & s \\ \hline
      s & t & t & s \\
      t & s & s & t \\
      0 & 0 & 0 & 0
    \end{array}
  \right] \,, \quad
  \mathbf{p}_{3} = \left[
    \begin{array}{cc|cc}
      0 & 0 & 0 & 0 \\
      t & s & t & s \\
      s & t & s & t \\ \hline
      s & t & t & s \\
      t & s & s & t \\
      0 & 0 & 0 & 0
    \end{array}
  \right] \,.
  \label{eq:P2andP3}
\end{equation}
The correlations \eref{eq:P2andP3} constitute ideal Tsirelson correlations (\eqnref{losstsirelson} with $\etal=1$ and re-labelled outcomes for Alice) and, therefore, are as non-local as the quantum mechanics allows. The important fact to notice is that for all $\mathbf{p}_{1}$, $\mathbf{p}_{2}$ and $\mathbf{p}_{3}$, if the no-click events are discarded, Eve has perfect knowledge on Alice's outcomes for $x=0$, as can be seen by inspection from the tables.

We consider the following mixture
\begin{equation}
\mathbf{p}^\lambda =  \lambda \mathbf{p}_{1} +  (1 - \lambda)
  \frac{1}{2}\left(\mathbf{p}_{2} + \mathbf{p}_{3}\right)
\end{equation}
and require the local efficiency to be outcome-independent. In particular, in order to reproduce the correlations \eref{losstsirelson}, we solve
\begin{equation}
  \etal = \lambda + (1 - \lambda) \tfrac{1}{2}
  = \lambda p + (1 - \lambda) \,
\end{equation}
for $\lambda$, so that $\mathbf{p}^\lambda=\mathbf{p}^\etal$. As a result, for $\lambda = \frac{1}{3 - 2 p} = \frac{5 + 2\sqrt{2}}{17}$ we obtain an attack for which the correlations \eref{losstsirelson} are recovered with local efficiency:
\begin{equation}
\etal = \frac{11 + \sqrt{2}}{17} \approx 0.73 \quad\left(> 1/\sqrt{2} \approx 0.707\right).
\label{eq.crit_Eric}
\end{equation}

As, once the non-conclusive events are discarded, Eve can then predict with certainty Alice's outcome for the setting $x=0$, the above attack invalidates the upper bound \eref{eq:GPSbound} that predicts this to be impossible for any $\etal>1/\sqrt{2}\approx 0.707$. Although the bound \eref{eq:GPSbound} cannot thus hold in complete generality for situations including losses (as already speculated in \citeref{curty2011}), it still remains to be proven whether \eqnref{eq:GPSbound} can be considered to be valid for the specific correlations arising in the protocol of~\citeref{gisin2010}.

\begin{figure*} [ht!]
\centering
\includegraphics[width=1\linewidth]{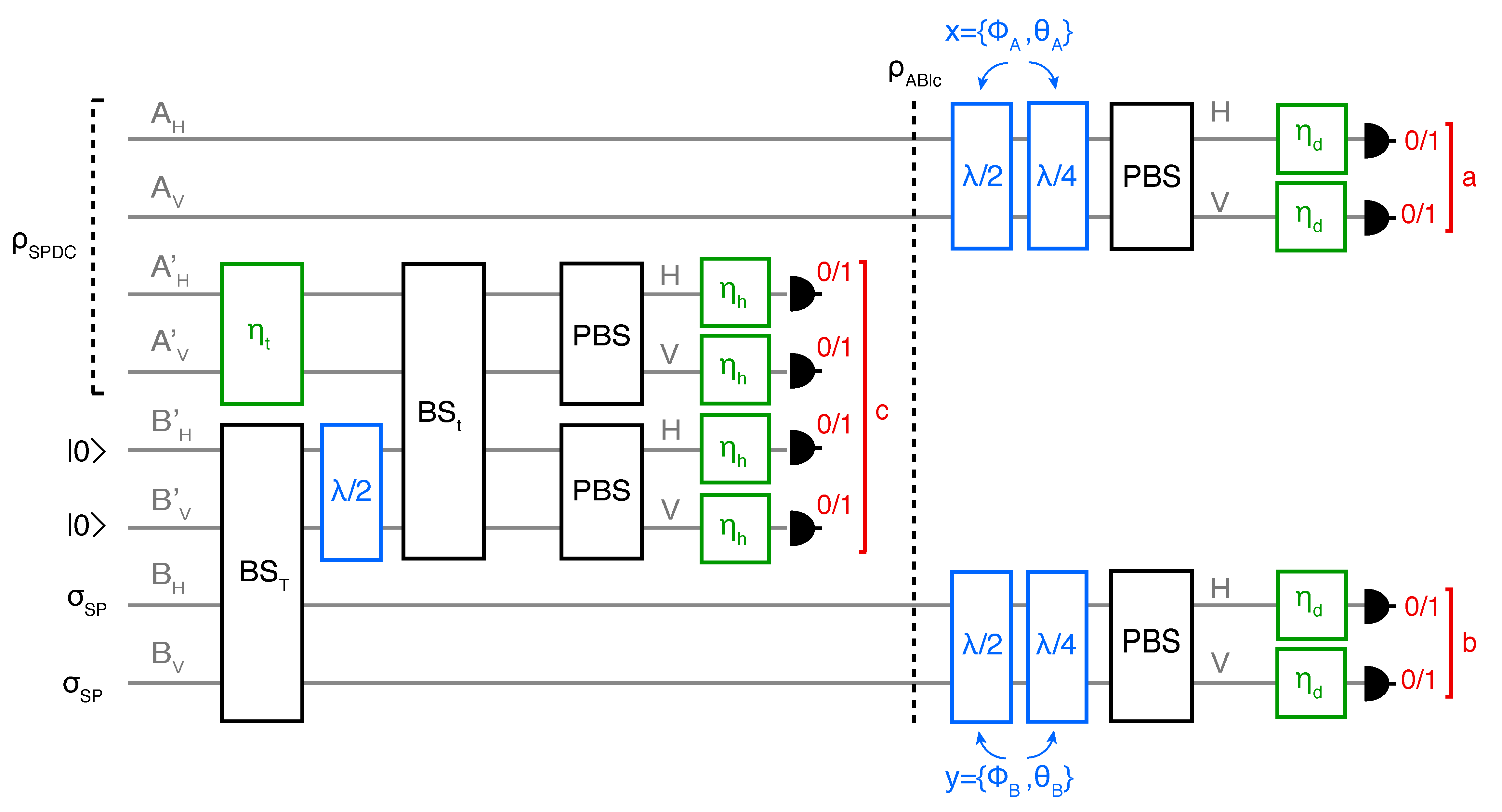}
\footnotesize{
\caption {\textbf{SH scheme of \figref{fig:scheme}(\textbf{a}) with all the photonic modes separated.} 
Alice employs a SPDC source to prepare entangled photon pairs in modes $\t{A}_{H/V}$ and $\t{A}'_{H/V}$ described by the state $\rho_\t{SPDC}$ in \eqnref{eq.spdc}. Bob uses two SP sources instead to simultaneously prepare single photons in modes $\t{B}_{H/V}$, each described by the state $\sigma_\t{SP}$ in \eqnref{eq.sp}. The whole SH scheme corresponds to a linear-optics circuit involving beamsplitters (BSs), half- and quarter-wave plates ($\frac{\lambda}{4}$ and $\frac{\lambda}{4}$), polarising beamsplitters (PBSs), and binary detectors yielding "0" (for no photons) or "1" (when one or more photons are detected). The finite efficiency of heralding detectors, as well as ones held by Alice and Bob, is accounted for by loss parameters $\eta_\t{h}$ and $\eta_\t{d}$, respectively, which similarly to the transmission loss, $\etat$, correspond to a BS-transformation with the vacuum state impinging the empty input port. In our analysis we consider the overall initial state $\rho^{(\text{SH})}_{\text{AB}}$ to be adequately described by its lowest-order expansion \eref{eq.SHinitial}. We propagate it then through the circuit in order to compute the resulting state shared by Alice and Bob conditioned on successful heralding outcome, i.e.~$\rho_{\text{AB}|c}^{(\mathrm{SH})}$ with $c=0110=:\herald$ when only the middle two of the heralding detectors click. $\rho_{\text{AB}|\herald}$ is then the state spanning modes $\t{A}_{H/V}$ and $\t{B}_{H/V}$ that Alice and Bob perform dual-rail polarisation qubit measurements on, whose settings are completely parametrised by the angles $x=\{\phi_\t{A},\theta_\t{A}\}$ and $y=\{\phi_\t{B},\theta_\t{B}\}$~\cite{ralph_chapter_2010}. Finally, the outcomes $a$ and $b$ correspond to the four possible click patterns observed by Alice and Bob, respectively. } 
\label{fig:SH_scheme}
}
\end{figure*}
%

%%%%%%%%%%%%%%%%%%%%%%%%%%%%%%%%%%%%%%%%%%%%%%%%%%%%%%%%%%%%%%%%%%%
\section{Heralded states produced by the SH and CH schemes} 
\label{app:states_CHSH}
In \figref{fig:SH_scheme} we depict once more the SH scheme---see its implementation in \figref{fig:scheme}(\textbf{a})---while separating explicitly the photonic modes involved, i.e.~distinct modes originating from the labs of Alice and Bob ($\t{A}$ and $\t{B}$) distinguished also by their polarisations ($H$ and $V$), as well as the auxiliary modes (labelled as primed "$'$") that effectively are the ones to reach the heralding station, and are measured to obtain the heralding signal $c$. A protocol round is then accepted within the SH (and also CH, see below) scheme only if the click pattern $c=0110=:\herald$ is observed with only the two middle detectors in \figref{fig:SH_scheme} clicking. 

Within the SH scheme Alice uses the SPDC process to produce a pair of entangled photons in modes $\t{A}_{H/V}$ and $\t{A}'_{H/V}$ described by the state \eref{eq.spdc}, $\rho_\t{SPDC}$. Bob employs on-demand sources in order to simultaneously prepare single photons (SPs) in modes $\t{B}_H$ and $\t{B}_V$, each described by the state \eqref{eq.sp}, $\sigma_\t{SP}$, see \figref{fig:SH_scheme}. Inspecting the expressions \eref{eq.spdc} and \eref{eq.sp}, the SH scheme ideally works at first order in $\bar{p}$ and zeroth order in $p$, respectively, with higher orders being negligible due to $\bar{p}\lesssim10^{-2}$~\cite{giustina2015,lynden2015} and $p\lesssim10^{-4}$~\cite{ding2016}. For completeness, however, we perform the analysis up to second order in both $p$ and $\bar{p}$ by considering the initial (unnormalised and uncorrelated) state of Alice and Bob---i.e.~the overall one present initially in the modes $\t{A}_{H/V}$, $\t{A}'_{H/V}$ and $\t{B}_{H/V}$ in \figref{fig:SH_scheme}---to read:
\begin{align}
\rho^{(\text{SH})}_{\text{AB}} & = 
 \ketbra{0^\t{AA'},1_H^\t{B}1_V^\t{B}}{...}+ p\ketbra{0^\t{AA'},1_H^\t{B}2_V^\t{B}}{...}+\nonumber\\
& p\ketbra{0^\t{AA'},2_H^\t{B}1_V^\t{B}}{...}+  p^2\ketbra{0^\t{AA'},2_H^\t{B}2_V^\t{B}}{...}   +\nonumber\\
& p^2\ketbra{0^\t{AA'},1_H^\t{B}3_V^\t{B}}{...} + p^2\ketbra{0^\t{AA'},3_H^\t{B}1_V^\t{B}}{...}+\nonumber\\
& \bar{p}\ketbra{\Psi_1^\t{AA'},1_H^\t{B}1_V^\t{B}}{...} +\nonumber\\
& p\bar{p}\ketbra{\Psi_1^\t{AA'},1_H^\t{B}2_V^\t{B}}{...} + p\bar{p}\ketbra{\Psi_1^\t{AA'},2_H^\t{B} 1_V^\t{B}}{...} +\nonumber\\
& \frac{3}{4}\bar{p}^2\ketbra{\Psi_2^\t{AA'},1_H^\t{B}1_V^\t{B}}{...} + \bigO{p^i\bar{p}^j}_{i+j=3}
\label{eq.SHinitial}
\end{align}
with higher-order terms yielding negligible contributions, which nonetheless must be later accounted for (see \appref{app:higher_terms}) when assuring the security of the DIQKD protocol.

Inspecting \eqnref{eq.SHinitial}, it is the seventh term occurring at $\bar{p}$-order which is the desired one, containing an entangled pair $\ket{\Psi_1}$ in modes $\t{A}_{H/V}$ and $\t{A}'_{H/V}$ and single photons in both modes $\t{B}_{H/V}$. All other terms are spurious:~the first six are associated with the vacuum production rounds of the SPDC source held by Alice; the eighth and ninth correspond to cases in which the SPDC process succeeds but one of the SPs emits two photons instead; while the last term appears due to double-pair production of the SPDC. In order to compute the state $\rho_{\text{AB}|c}^{(\mathrm{SH})}$ marked in \figref{fig:SH_scheme}, we propagate the initial state \eref{eq.SHinitial} ``term by term'' through the relevant parts of the circuit and account for the photon-detection measurement in modes $\t{A}'_{H/V}$ and $\t{B}'_{H/V}$, while assuming the detectors to be binary (on/off), i.e.~clicking with efficiency $\eta_\t{h}$,  without distinguishing the exact photon number.

Although we omit here the explicit expression for $\rho_{\text{AB}|c}^{(\mathrm{SH})}$ that we obtain for $c=\herald$ (i.e.~when only two out of the four relevant detectors in \figref{fig:SH_scheme} click), we note that the leading order of $\rho_{\text{AB}|\herald}^{(\mathrm{SH})}$ stated in \eqnref{eq.SHstate} of the main text is the result of the desired contribution---the seventh term in \eqnref{eq.SHinitial}. Crucially, contributions of all the other terms in \eqnref{eq.SHinitial} are suppressed due to $p,\bar{p}\ll1$. Moreover, the impact of finite detection efficiency, $\eta_\t{h}<1$, can be assumed to affect only the probability of successful heralding, i.e.~$P(c\!=\!\herald)$ that corresponds to the norm of $\rho_{\text{AB}|\herald}^{(\mathrm{SH})}$, because we ensure that $T\approx1$ within the SH scheme. As a result, the SPs produced by Bob hardly enter the modes $\t{B}'_{H/V}$ in \figref{fig:SH_scheme} or, in other words, leave the lab of Bob in \figref{fig:scheme}(\textbf{a}). 

Importantly, we use the full expression for $\rho_{\text{AB}|\herald}^{(\mathrm{SH})}$ incorporating all the contributions of \eqnref{eq.SHinitial} to compute the resulting correlations shared by Alice and Bob after they measure photons in modes $\t{A}_{H/V}$ and $\t{B}_{H/V}$ in \figref{fig:SH_scheme}, respectively, i.e.:
\begin{equation}
P^\t{(SH)}\!\left(a,b\left|\!
\begin{array}{c}
x\!=\!\{\phi_\t{A},\theta_\t{A}\},y\!=\!\{\phi_\t{B},\theta_\t{B}\},t,\eta_\t{d},c\!=\!\herald\\
\eta_\t{h},\etat,T\approx1,p\approx0,\bar{p}\approx0
\end{array}\!\!
\right.\right)\!, 
\label{eq:joint_prob_SH}
\end{equation}
which similarly to the initial state \eref{eq.SHinitial} is valid up to $\bigO{p^i\bar{p}^j}_{i+j=3}$. The form of the joint probability distribution \eref{eq:joint_prob_SH} depends strongly on the measurement settings controlled by the angles ($\phi$, $\theta$) of polarization dual-rail qubits~\cite{ralph_chapter_2010} detected by Alice and Bob, the efficiency $\eta_\t{d}$ of the binary detectors they employ, as well as the $t$-parameter controlling the asymmetry of the heralding BSM (see \figref{fig:SH_scheme}) and, hence, the partial entanglement of the target state $\ket{\psi_\t{AB}^t}$ in \eqnref{eq.SHstate}. Nonetheless, we also list in the second row in \eqnref{eq:joint_prob_SH} all the other parameters that the shared correlations formally depend on due to higher-order terms taken into account within the initial state \eref{eq.SHinitial}. 

However, in practice---as verified also by our numerical analysis---the dependence on the transmission loss parameter, $\etat$, as well as the efficiency of heralding detectors, $\eta_\t{h}$, can be completely disregarded as they enter \eqnref{eq:joint_prob_SH} at higher order in $p$ and $1-T$, respectively. Nonetheless, let us emphasise that to compute both the 
critical local efficiencies, $\etalC$, stated in \tabref{tab.1} and the DIQKD key rates presented in \figref{fig.secretrate}, we use the full expression for the joint probability \eref{eq:joint_prob_SH}. In particular, we set the efficiency of the heralding detectors to be equal to the ones of Alice and Bob, i.e.~$\eta_\t{h}=\eta_\t{d}=:\etal$, which in practice affects then only the key rate with $P(c\!=\!\herald)\propto \eta_\t{h}^2$ in \eqnref{keyDIQKD}. Moreover, in order to determine the highest key rates $K$ in \eqnref{keyDIQKD} that yield the lowest critical local efficiency, $\etalC=\eta_\t{d}^*$, we also fine-tune the source parameters $\{\bar{p},p,T\}$, whose orders of magnitude we importantly constrain to $\bar{p}\approx10^{-2}$, $p\approx10^{-4}$ and $(1-T)\approx10^{-3}$ for the lowest-order expansion analysis to always be valid.

For the CH scheme depicted \figref{fig:scheme}(\textbf{b}), we follow exactly the same analysis as stated above for the SH scheme. The CH scheme can be presented as a similar linear optics circuit, where now the A-modes constitute just a copy (mirror image) of the B-modes drawn in \figref{fig:SH_scheme}. Within the CH scheme both Alice and Bob possess two on-demand SP sources. The only difference---due to the heralding station in \figref{fig:scheme}(\textbf{b}) being held outside of the labs---are the transmission losses, $\etat$, that must now be accounted for not only in the $\t{A}'_{H/V}$ (see \figref{fig:SH_scheme}) but also in the $\t{B}'_{H/V}$ modes. 

However, for the CH scheme the initial state prepared by Alice and Bob (this time using only the four modes $\t{A}_{H/V}$ and $\t{B}_{H/V}$ in \figref{fig:SH_scheme} with others containing vacuum) no longer contains spurious vacuum contributions, due to the SPDC process being absent, i.e.:
\begin{align}
\rho^{(\text{CH})}_{\text{AB}} = 
& \ketbra{1_H^A1_V^A,1_H^B1_V^B}{...}+ p\ketbra{1_H^A1_V^A,1_H^B2_V^B}{...}+ \nonumber \\
& p\ketbra{1_H^A1_V^A,2_H^B1_V^B}{...}   + p\ketbra{1_H^A2_V^A,1_H^B1_V^B}{...}+\nonumber\\
& p\ketbra{2_H^A1_V^A,1_H^B1_V^B}{...}  + \bigO{p^2},
\label{eq.CHinitial}
\end{align}
where, in contrast to \eqnref{eq.CHinitial}, the ideal contribution occurs at zeroth order in $p$, that is, when each of the four SPs produces a single photon. Still, similarly to the SH scheme, we include higher-order contributions (now, at first order in $p$) in our analysis. 

In particular, we compute the corresponding state $\rho^{(\text{CH})}_{\text{AB}|\herald}$ conditioned on successful heralding, whose main contribution comes from the zeroth-order in \eqnref{eq.CHinitial} stated in \eqnref{eq.CHstate} of the main text. As in \eqnref{eq:joint_prob_SH}, while keeping all the contributions of the initial state \eref{eq.CHinitial}, we compute the shared correlations of Alice and Bob after they perform their measurements, i.e.: 
\begin{equation}
P^\t{(CH)}\!\left(a,b\left|\!
\begin{array}{c}
x\!=\!\{\phi_\t{A},\theta_\t{A}\},y\!=\!\{\phi_\t{B},\theta_\t{B}\},t,\eta_\t{d},c\!=\!\herald\\
\tilde{\etat}=\eta_\t{h}\etat,T\approx0,p\approx0
\end{array}\!\!
\right.\right)\!, 
\label{eq:joint_prob_CH}
\end{equation}
where in contrast to \eqnref{eq:joint_prob_SH} the efficiency of the heralding detectors $\eta_\t{h}$---appearing in \eqnref{eq:joint_prob_CH} again only due to higher-order terms in \eqnref{eq.CHinitial}---can be interpreted as just another source of effective transmission loss, $\tilde{\etat}$. As a result, in order for comparison of the key rates in \figref{fig.secretrate} between the SH and CH schemes to be fair, we rescale $\etat\to\etat\etal$ in case of the latter to account for the heralding detectors to have the same efficiency as the ones held by Alice and Bob (i.e., $\eta_\t{h}=\eta_\t{d}=\etal$ in \figref{fig:SH_scheme}). Otherwise, we perform exactly the same analysis for the joint distribution \eref{eq:joint_prob_CH} as for \eqnref{eq:joint_prob_SH}, in order to determine the maximal key rate, $K$ in \eqnref{fig.secretrate}, and critical local efficiencies, $\etalC$ in \tabref{tab.1}, where we ensure now that $T\approx10^{-3}$ and $p\lessapprox10^{-3}$ throughout the numerical optimisation, so that our perturbative approach (in photon number) assumed by \eqnref{eq.CHinitial} always holds.

Finally, let us note that within both the SH and CH schemes we may naturally account for the finite efficiency of the on-demand sources employed~\cite{aharonovich2016}, which produce the SPs in the state \eqref{eq.sp}, i.e.~$\sigma_\t{SP}$ marked in \figref{fig:SH_scheme} for the SH scheme. Inspecting \figref{fig:SH_scheme} and, in particular, modes $\t{B}_{H/V}$---and similarly for the $\t{A}_{H/V}$ modes in case of the CH scheme, in which they are equivalent---it becomes clear that one may propagate beam-splitters responsible for the finite detection, $\eta_\t{d}$, all the way through the circuit onto the initial state without altering the scheme on the whole. Hence, given that each SP-source works with $\eta_\t{s}$-efficiency, all our analysis applies with now simply the overall local efficiency reading $\etal=\eta_\t{s}\eta_\t{d}$, so that it accounts for the finite efficiency of both the sources and detectors contained within the lab of Alice or Bob, or both (as summarised in the main text while including also finite transmission between these components, $\eta_\t{lt}$).

%%%%%%%%%%%%%%%%%%%%%%%%%%%%%%%%%%%%%%%%%%%%%%%%%%%%%%%%%%%%%%%%%%%
\section{Guessing probability} 
\label{app:pguess}
The min-entropy term $-\log_2 G_{\textbf{p}}(x^*)$ in \eqnref{eq.secret} of the main text is expressed with help of the \textit{device-independent guessing probability}, i.e.~the average probability that the eavesdropper Eve correctly guesses the output of Alice using an optimal strategy:~\cite{nieto2014}
\begin{eqnarray}
\label{eq.Pguess}
G_{\textbf{p}}(x^*)\
&:=&\
\underset{\{\textbf{p}^e\}}{\max}\ \ \sum\limits_{\substack{e}} P(e)\,P(a=e|x^*,e)\\
&&\ \ \text{s.t.}\quad
\sum\limits_{\substack{e}} \textbf{p}^{e} = \textbf{p} \;\;\text{and}\;\;\forall e:\,\textbf{p}^{e} \in \widetilde{Q}.\nonumber
\end{eqnarray}
Here, $P(e)$ denotes the probability that Eve observes the outcome $e$, while $P(a=e|x^*,e)$ effectively represents the probability that Alice obtains an outcome $a$ coinciding with $e$, given to be the one observed by Eve. 

Any strategy of Eve in \eqnref{eq.Pguess} can be seen as a measurement that she performs on her system, which then produces a decomposition (a collection) of unnormalized behaviours $\{\textbf{p}^{e}\}$ distributed between Alice and Bob. The guessing probability \eref{eq.Pguess} is then obtained by maximising the success of Eve's strategy over all such possible decompositions that, however, must reproduce on average the behaviour $\textbf{p}$ observed by Alice and Bob and be compatible with quantum mechanics (see the second line of \eqnref{eq.Pguess}). Formally, each of them must belong to the set of unnormalised behaviours $\widetilde{Q}$ which stem from the Born's rule when valid quantum measurements act on an unnormalized, yet unspecified, quantum state. Thus, to enforce the \textit{quantumness} of Eve's strategy, the second constraint in \eqnref{eq.Pguess} demands that all $\textbf{p}^{e}$ belong to  $\widetilde{Q}$.

Imposing membership in $\widetilde{Q}$ is difficult since a precise characterization of $\widetilde{Q}$ is unknown. However, semi-definite programming (SDP) relaxations similar to the ones presented by \citet{navascues2007} can be introduced to bound $G_{\textbf{p}}(x^*)$ from above~\cite{nieto2014}. One defines a convergent hierarchy of convex sets that have a precise characterization and obey $\widetilde{Q}_1 \supseteq \widetilde{Q}_2 \supseteq ... \supseteq  \widetilde{Q} $. This hierarchy approximates the quantum set $\widetilde{Q}$ from outside, so that any optimisation over the quantum set can be relaxed (to some order $k$) by replacing $\widetilde{Q}$ in \eqnref{eq.Pguess} with $\widetilde{Q}_k$. Hence, the program presented in \eqnref{eq.Pguess} becomes an SDP when relaxations of the set $\widetilde{Q}$ are employed---in our work we mostly consider relaxations to the order $1+AB$, i.e, an intermediary order between first and second orders. 

Finally, let us note that from the dual formulation~\cite{boyd2004} of the SDP program employed, we are also always able to retrieve the Bell inequality that is optimal for bounding the degree of predictability that a quantum eavesdropper may have about the string of Alice's outcomes~\cite{nieto2014}.

%%%%%%%%%%%%%%%%%%%%%%%%%%%%%%%%%%%%%%%%%%%%%%%%%%%%%%%%%%%%%%%%%%%
\section{Dealing with higher-order multi-photon contributions} 
\label{app:higher_terms}
In order to deal with quantum states produced by SPDC and single-photon sources (presented in \appref{app:states_CHSH}), one typically truncates the global state produced by all sources in the setup up to a certain order $n$~\cite{gisin2010,curty2011,meyerscott2013}. Since any setup we consider is powered by SPDC sources parameterized by $\bar{p}$ and single-photon sources parametrized by $p$, a truncation to the order, e.g.~$n=2$ of the global state---which is the tensor product of the states of each source---would keep all terms up to order $\bigO{p^2}$, $\bigO{p \bar{p}}$ and  $\bigO{\bar{p}^2}$.

Nevertheless, this perturbative approximation may yield misleading conclusions about the nonlocal character of the observed correlations and compromise DIQKD security for a given setup. In fact, one has to guarantee that contributions not considered in the truncation will not contradict the conclusions about the nonlocal character of the behaviour in question.

To avoid this problem, we develop here a method based on SDP techniques where all high-order contributions ($>\!n$) that are not taken into account are fully controlled by Eve, to her benefit. This may seem too conservative, but the method turns out to be efficient and not overly pessimistic, since the contribution of high-order terms becomes irrelevant for sufficiently low values of $p$ and $\bar{p}$.

The key idea is to conceive higher-order contributions as producing an unknown and uncharacterized quantum behaviour $\textbf{p}_Q$ prepared by Eve for Alice and Bob. If $\textbf{p}^{\text{est}}_{n}$ denotes the estimation of the behaviour of Alice and Bob constructed to the order $n$ (e.g.~one derived basing on states \eref{eq.SHinitial} or \eref{eq.CHinitial} for SH and CH-schemes, respectively), then the first step of the method is to write the observed behaviour $\textbf{p}$ as a convex decomposition: $\textbf{p} = (1-\epsilon_n)\textbf{p}^{\text{est}}_{n} + \epsilon_n\textbf{p}_Q$.  

At the quantum level, the total state being shared, given a collection of sources producing a perturbative state such as \eref{eq.spdc}, may be written as a convex mixture $p(n)\rho_n + p(\bar{n})\rho_{\bar{n}}$, where $\rho_n$ is the truncated state according to the estimation made at some order $n$. $\rho_{\bar{n}}$ is thus the remaining ``tail'' of high-order contributions, and $p(\bar{n})=1-p(n)$. Moving to the level of probability distributions, linearity of Born's rule with respect to $\rho$ implies that the elements of the observed behaviour $\textbf{p}$ conditioned on the outcome $c$ employed in the heralding stage (see \appref{app:states_CHSH}) may be decomposed in a similar fashion, i.e.:
\begin{equation}
P(a,b|c) =  p(n|c)\,P(a,b|c,n) + p(\bar n|c)\,P(a,b|c,\bar{n}).
\end{equation}
The probabilities $P(a,b|c,n)$ above are then nothing but the elements of the estimated behaviour $\textbf{p}_n^{\text{est}}$ computed up to the $n$th order. 

We rewrite $p(n|c)$ employing the Bayes rule:
\begin{equation}
p(n|c)=\frac{p(c|n)p(n)}{p(c)}.
\label{eq.bayes}
\end{equation}
The numerator in \eqnref{eq.bayes} is known, as $p(c|n)$ is merely the probability of observing the heralding outcome $c$ while assuming the $n$th order truncation at the level of the sources. The denominator, however, is unknown and corresponds to the probability of observing the outcome $c$, without assuming any truncation. 

Still, it is possible to set an upper bound on $p(c)$:
\begin{align}
p(c) 
& =\sum_{\vec{k}=\vec{0}}^{\infty}p(\vec{k})p(c|\vec{k})  \nonumber \\
& \leq\; p_{\vec{K}_n}(c):=\sum_{\vec{k}=\vec{0}}^{\vec{K}_n}p(\vec{k})p(c|\vec{k}) + \sum_{\vec{k}>\vec{K}_n}^{\infty}p(\vec{k}),
\label{eq.app.pc}
\end{align}
where the vector of variables $\vec{k}=(k_1,k_2,...,k_s)$ describes the possible number of photons produced by each of the $s$ sources. In particular, $p(\vec{k})$ gives the distribution for each of the possible combinations of photons (or photon pairs) occurring, when produced by the sources. Vector $\vec{K}_n$ contains the numbers of photons that each source can maximally produce, given a particular order $n$ of the truncation.  

Bounding \eqnref{eq.bayes} with help of \eqnref{eq.app.pc}, one gets the desired upper bound on $\epsilon_n=1-p(n|c)$, i.e.,
\begin{equation}
\epsilon_n 
\;\le\; 
\epsilon_n^\uparrow:=
1-\frac{p(c|n)p(n)}{p_{\vec{K}_n}(c)}, 
\label{eq.app.eps}
\end{equation}
which can be importantly computed for a given optical scheme and the order $n$ assumed. Consistently, $\epsilon_n^\uparrow$ (and, hence, $\epsilon_n$) goes to zero as the order $n$ increases, so that $\textbf{p}^{\text{est}}_{n}$ converges to $\textbf{p}$ in the limit $n\to\infty$.

In an analogous way to \eqnref{eq.Pguess}, we define the \emph{device-independent guessing probability to the order} $n$ as:
\begin{align}
G_{\textbf{p}_n^\text{est}}(x^*, \epsilon):=\underset{\{\textbf{p}^e\}}{\max}
&\, 
\sum\limits_{\substack{e}}P(e,a=e|x^*)
\label{eq.Pguessn}\\
\text{s.t.}
&\,
\sum\limits_{\substack{e}} \textbf{p}^{e} = (1-\epsilon)\textbf{p}^{\text{est}}_{n} + \epsilon\textbf{p}_Q, \nonumber \\
&\;\;
\textbf{p}_Q \in Q \;\;\text{and}\;\;\forall e:\,\textbf{p}^{e} \in \widetilde{Q}, \nonumber
\end{align}
where ($\widetilde{Q}$)$Q$  denotes the set of (un)normalized quantum behaviours. The crucial difference between \eqnsref{eq.Pguess}{eq.Pguessn} is that Eve is now \textit{not} obliged to reproduce exactly the behaviour $\textbf{p}$ with her collection of unnormalised behaviours $\{\textbf{p}^{e}\}$.  Instead, she possesses a supplementary quantum behaviour $\textbf{p}_Q $ that she can tailor, so that it is easier for her to reproduce the behaviour $\textbf{p}^{\text{est}}_{n}$ for a given \emph{fixed} value of $\epsilon$ and, hence, better guess the outcome of Alice's box.

Now, the following inequalities must hold: 
\begin{equation}
G_{\textbf{p}_n^\text{est}}(x^*)
\;\le\;
G_{\textbf{p}_n^\text{est}}(x^*, \epsilon_n)
\;\le\;
G_{\textbf{p}_n^\text{est}}(x^*, \epsilon_n^\uparrow),
\end{equation}
where the first one is guaranteed to be saturated whenever it is optimal to set $\textbf{p}_Q=\textbf{p}^{\text{est}}_{n}$ in \eqnref{eq.Pguessn} (i.e.~the truncation plays no role), while the second one whenever the bound \eref{eq.app.eps} is tight.
As a result, we may always upper-bound Eve's optimal guessing probability by $G_{\textbf{p}_n^\text{est}}(x^*, \epsilon_n^\uparrow)$, and by doing so we can only underestimate the attainable key rate of the DIQKD protocol---see \eqnref{eq.secret} of the main text.

Let us stress that the method presented above is quite general, as it can be applied to any other uncharacterised imperfection parametrised by $\epsilon$, such that its action arises as convex decomposition of the form $\textbf{p} = (1-\epsilon)\textbf{p}^{\text{est}} + \epsilon\textbf{p}_Q$. In particular, it allows to upper-bound the guessing probability for any type of noise that may be represented as a convex mixture at the level of a quantum state, given that the corresponding mixing probability can also be bound from above by a known $\epsilon^\uparrow<1$.

%%%%%%%%%%%%%%%%%%%%%%%%%%%%%%%%%%%%%%%%%%%%%%%%%%%%%%%%%%%%%%%%%%%
\section{Noise robustness for nonlocality} 
\label{app:robust}

We analyze the robustness to white noise of the estimated behaviours $\textbf{p}^{\text{est}}$ that our two schemes produce. We determine the maximal value $w^*$ of white noise $\mathds{1}_{\textbf{p}}$---a distribution in which all the outcomes are equally likely, independently of the measurement choices---which can be convexly added such that the behaviour $(1-w)\textbf{p}^{\text{est}} + w\mathds{1}_{\textbf{p}}$ remains nonlocal.

Membership of a probability distribution to the set of local behaviours~\cite{brunner2014} is an instance of a linear program~\cite{boyd2004}. Geometrically speaking, the set of local behaviours is a polytope in the space of probability distributions, whose extremal points correspond to %a finite number of fixed
particular deterministic strategies $\{\textbf{D}_{\mu}\}_{\mu}$ that are sufficient to decompose any local behaviour. In fact, there is a finite number of such deterministic strategies, and the white noise tolerance of $\textbf{p}^{\text{est}}$ is given by the solution of the following linear program:
\begin{eqnarray}
\label{eq.Wnoise}
w^*\
&=&\
\underset{\{q_{\mu}\}}{\min}\ \ w\\
&&\ \ \text{s.t.}\;
\ (1-w)\textbf{p}^{\text{est}} + w\mathds{1}_{\textbf{p}} = \sum_{\mu}q_{\mu}\textbf{D}_{\mu}, \nonumber \\
&&\quad
\ \ \ \ \  \sum_{\mu}q_{\mu}=1\;\;\text{and}\;\; \forall \mu:\,q_{\mu}\geq 0.\nonumber
\end{eqnarray}
The white-noise tolerance threshold, $w^*$, should be interpreted as deviations from the desired correlations at the level of probability distributions. This is the worst-case approach in which the experimental imperfections not accounted for in $\textbf{p}^{\text{est}}$ provide Alice and Bob with completely uncorrelated results. The fact that our schemes tolerate high amounts of white noise (see \tabref{tab.1} of the main text) ensures that our results will not be strongly affected when introducing other sources of noise, not accounted for in the analysis.

\bibliographystyle{unsrtnat}
\bibliography{diqkd}

\end{document}